\documentclass[journal,twocolumn]{IEEEtran}
\usepackage{epsfig,algorithm,amsmath,amssymb,color,subfigure,url}

\author{
\authorblockN{Namrata Vaswani \\
}
\authorblockA{
Dept. of Electrical and Computer Engineering \\
Iowa State University, Ames, IA 50011, USA \\
Email: namrata@iastate.edu
}
\thanks{This work was supported by NSF grants ECCS-0725849 and CCF-0917015. A part of this work was presented at Allerton 2010 \cite{stability_allerton}.}
}

\title{Stability of Modified-CS and LS-CS for Recursive Reconstruction of Sparse Signal Sequences} 

\begin{document}
\setlength{\arraycolsep}{0.03cm}
\newcommand{\xhat}{\hat{x}}
\newcommand{\xpred}{\hat{x}_{t|t-1}}
\newcommand{\Ppred}{P_{t|t-1}}
\newcommand{\ty}{\tilde{y}_t}
\newcommand{\tty}{\tilde{y}_{t,\text{res}}}
\newcommand{\tw}{\tilde{w}_t}
\newcommand{\ttw}{\tilde{w}_{t,f}}
\newcommand{\betahat}{\hat{\beta}}

\newcommand{\ypast}{y_{1:t-1}}
\newcommand{\sone}{S_{*}}
\newcommand{\sinf}{{S_{**}}}
\newcommand{\smax}{S_{\max}}
\newcommand{\smin}{S_{\min}}
\newcommand{\samax}{S_{a,\max}}
\newcommand{\Nhat}{{\hat{N}}}

\newcommand{\sgn}{\text{sgn}}

\newcommand{\Dnum}{D_{num}}
\newcommand{\pss}{p^{**,i}}
\newcommand{\fr}{f_{r}^i}

\newcommand{\A}{{\cal A}}
\newcommand{\Z}{{\cal Z}}
\newcommand{\B}{{\cal B}}
\newcommand{\R}{{\cal R}}
\newcommand{\reg}{{\cal G}}
\newcommand{\const}{\mbox{const}}

\newcommand{\trace}{\mbox{tr}}

\newcommand{\hsim}{{\hspace{0.0cm} \sim  \hspace{0.0cm}}}
\newcommand{\he}{{\hspace{0.0cm} =  \hspace{0.0cm}}}

\newcommand{\vect}[2]{\left[\begin{array}{cccccc}
     #1 \\
     #2
   \end{array}
  \right]
  }

\newcommand{\matr}[2]{ \left[\begin{array}{cc}
     #1 \\
     #2
   \end{array}
  \right]
  }
\newcommand{\vc}[2]{\left[\begin{array}{c}
     #1 \\
     #2
   \end{array}
  \right]
  }

\newcommand{\gdot}{\dot{g}}
\newcommand{\Cdot}{\dot{C}}
\newcommand{\re}{\mathbb{R}}
\newcommand{\n}{{\cal N}}  
\newcommand{\N}{{\overrightarrow{\bf N}}}  
\newcommand{\chat}{\tilde{C}_t}
\newcommand{\chati}{\chat^i}

\newcommand{\cmin}{C^*_{min}}
\newcommand{\twi}{\tilde{w}_t^{(i)}}
\newcommand{\twj}{\tilde{w}_t^{(j)}}
\newcommand{\wi}{{w}_t^{(i)}}
\newcommand{\twio}{\tilde{w}_{t-1}^{(i)}}

\newcommand{\tWi}{\tilde{W}_n^{(m)}}
\newcommand{\tWj}{\tilde{W}_n^{(k)}}
\newcommand{\Wi}{{W}_n^{(m)}}
\newcommand{\tWio}{\tilde{W}_{n-1}^{(m)}}

\newcommand{\ds}{\displaystyle}

\newcommand{\SAR}{S$\!$A$\!$R }
\newcommand{\MAR}{MAR}
\newcommand{\MMRF}{MMRF}
\newcommand{\AR}{A$\!$R }
\newcommand{\GMRF}{G$\!$M$\!$R$\!$F }
\newcommand{\DTM}{D$\!$T$\!$M }
\newcommand{\MSE}{M$\!$S$\!$E }
\newcommand{\RCS}{R$\!$C$\!$S }
\newcommand{\uomega}{\underline{\omega}}
\newcommand{\y}{v}
\newcommand{\x}{w}
\newcommand{\lu}{\mu}
\newcommand{\g}{g}
\newcommand{\bft}{{\bf t}}
\newcommand{\refmap}{{\cal R}}
\newcommand{\totrefl}{{\cal E}}
\newcommand{\beq}{\begin{equation}}
\newcommand{\eeq}{\end{equation}}
\newcommand{\bdm}{\begin{displaymath}}
\newcommand{\edm}{\end{displaymath}}
\newcommand{\hatz}{\hat{z}}
\newcommand{\hatu}{\hat{u}}
\newcommand{\tilz}{\tilde{z}}
\newcommand{\tilu}{\tilde{u}}
\newcommand{\hhatz}{\hat{\hat{z}}}
\newcommand{\hhatu}{\hat{\hat{u}}}
\newcommand{\tilc}{\tilde{C}}
\newcommand{\hatc}{\hat{C}}
\newcommand{\tim}{n}

\newcommand{\ssp}{\renewcommand{\baselinestretch}{1.0}}
\newcommand{\defd}{\mbox{$\stackrel{\mbox{$\triangle$}}{=}$}}
\newcommand{\goes}{\rightarrow}
\newcommand{\tends}{\rightarrow}
\newcommand{\defn}{\triangleq} 
\newcommand{\se}{&=&}
\newcommand{\sdefn}{& \defn  &}
\newcommand{\sle}{& \le &}
\newcommand{\sge}{& \ge &}
\newcommand{\plusminus}{\stackrel{+}{-}}
\newcommand{\Ey}{E_{Y_{1:t}}}
\newcommand{\ey}{E_{Y_{1:t}}}

\newcommand{\equivto}{\mbox{~~~which is equivalent to~~~}}
\newcommand{\nonzero}{i:\pi^n(x^{(i)})>0}
\newcommand{\nonzeroc}{i:c(x^{(i)})>0}

\newcommand{\supn}{\sup_{\phi:\|\phi\|_\infty \le 1}}

\newtheorem{theorem}{Theorem}
\newtheorem{lemma}{Lemma}
\newtheorem{corollary}{Corollary}
\newtheorem{definition}{Definition}
\newtheorem{remark}{Remark}
\newtheorem{example}{Example}
\newtheorem{ass}{Assumption}
\newtheorem{proposition}{Proposition}

\newtheorem{fact}{Fact}
\newtheorem{heuristic}{Heuristic}
\newcommand{\eps}{\epsilon}
\newcommand{\bd}{\begin{definition}}
\newcommand{\ed}{\end{definition}}
\newcommand{\udq}{\underline{D_Q}}
\newcommand{\td}{\tilde{D}}
\newcommand{\epsinv}{\epsilon_{inv}}
\newcommand{\al}{\mathcal{A}}

\newcommand{\bfx} {\bf X}
\newcommand{\bfy} {\bf Y}
\newcommand{\bfz} {\bf Z}
\newcommand{\ddas}{\mbox{${d_1}^2({\bf X})$}}
\newcommand{\ddbs}{\mbox{${d_2}^2({\bfx})$}}
\newcommand{\dda}{\mbox{$d_1(\bfx)$}}
\newcommand{\ddb}{\mbox{$d_2(\bfx)$}}
\newcommand{\xinc}{{\bfx} \in \mbox{$C_1$}}
\newcommand{\eqa}{\stackrel{(a)}{=}}
\newcommand{\eqb}{\stackrel{(b)}{=}}
\newcommand{\eqe}{\stackrel{(e)}{=}}
\newcommand{\leqc}{\stackrel{(c)}{\le}}
\newcommand{\leqd}{\stackrel{(d)}{\le}}

\newcommand{\leqa}{\stackrel{(a)}{\le}}
\newcommand{\leqb}{\stackrel{(b)}{\le}}
\newcommand{\leqe}{\stackrel{(e)}{\le}}
\newcommand{\leqf}{\stackrel{(f)}{\le}}
\newcommand{\leqg}{\stackrel{(g)}{\le}}
\newcommand{\leqh}{\stackrel{(h)}{\le}}
\newcommand{\leqi}{\stackrel{(i)}{\le}}
\newcommand{\leqj}{\stackrel{(j)}{\le}}

\newcommand{\w}{{W^{LDA}}}
\newcommand{\halpha}{\hat{\alpha}}
\newcommand{\hsigma}{\hat{\sigma}}
\newcommand{\slmax}{\sqrt{\lambda_{max}}}
\newcommand{\slmin}{\sqrt{\lambda_{min}}}
\newcommand{\lmax}{\lambda_{max}}
\newcommand{\lmin}{\lambda_{min}}

\newcommand{\da} {\frac{\alpha}{\sigma}}
\newcommand{\chka} {\frac{\check{\alpha}}{\check{\sigma}}}
\newcommand{\sumo}{\sum _{\underline{\omega} \in \Omega}}
\newcommand{\distance}{d\{(\hatz _x, \hatz _y),(\tilz _x, \tilz _y)\}}
\newcommand{\col}{{\rm col}}
\newcommand{\rcs}{\sigma_0}
\newcommand{\CalR}{{\cal R}}
\newcommand{\df}{{\delta p}}
\newcommand{\dq}{{\delta q}}
\newcommand{\dZ}{{\delta Z}}
\newcommand{\pprime}{{\prime\prime}}

\newcommand{\vn}{N}

\newcommand{\bv}{\begin{vugraph}}
\newcommand{\ev}{\end{vugraph}}
\newcommand{\bi}{\begin{itemize}}
\newcommand{\ei}{\end{itemize}}
\newcommand{\ben}{\begin{enumerate}}
\newcommand{\een}{\end{enumerate}}
\newcommand{\be}{\protect\[}
\newcommand{\ee}{\protect\]}
\newcommand{\bean}{\begin{eqnarray*} }
\newcommand{\eean}{\end{eqnarray*} }
\newcommand{\bea}{\begin{eqnarray} }
\newcommand{\eea}{\end{eqnarray} }
\newcommand{\nn}{\nonumber}
\newcommand{\ba}{\begin{array} }
\newcommand{\ea}{\end{array} }
\newcommand{\ep}{\mbox{\boldmath $\epsilon$}}
\newcommand{\epp}{\mbox{\boldmath $\epsilon '$}}
\newcommand{\Lep}{\mbox{\LARGE $\epsilon_2$}}
\newcommand{\und}{\underline}
\newcommand{\pdif}[2]{\frac{\partial #1}{\partial #2}}
\newcommand{\odif}[2]{\frac{d #1}{d #2}}
\newcommand{\dt}[1]{\pdif{#1}{t}}
\newcommand{\urho}{\underline{\rho}}

\newcommand{\spc}{{\cal S}}
\newcommand{\tspc}{{\cal TS}}

\newcommand{\uv}{\underline{v}}
\newcommand{\us}{\underline{s}}
\newcommand{\uc}{\underline{c}}
\newcommand{\utheta}{\underline{\theta}^*}
\newcommand{\ualpha}{\underline{\alpha^*}}

\newcommand{\uxy}{\underline{x}^*}
\newcommand{\uxyj}{[x^{*}_j,y^{*}_j]}
\newcommand{\arcl}[1]{arclen(#1)}
\newcommand{\one}{{\mathbf{1}}}

\newcommand{\uxyjt}{\uxy_{j,t}}
\newcommand{\E}{\mathbb{E}}

\newcommand{\rhomat}{\left[\begin{array}{c}
                        \rho_3 \ \rho_4 \\
                        \rho_5 \ \rho_6
                        \end{array}
                   \right]}
\newcommand{\deltat}{\tau} 
\newcommand{\deltatt}{\Delta t_1}
\newcommand{\ceil}[1]{\ulcorner #1 \urcorner}

\newcommand{\xxi}{x^{(i)}}
\newcommand{\txi}{\tilde{x}^{(i)}}
\newcommand{\txj}{\tilde{x}^{(j)}}

\newcommand{\mi}[1]{{#1}^{(m,i)}}

\setlength{\arraycolsep}{0.05cm}
\newcommand{\rest}{{T_\text{rest}}}
\newcommand{\zetahat}{\hat{\zeta}}
\newcommand{\tDelta}{{\tilde{\Delta}}}
\newcommand{\tDeltae}{{\tilde{\Delta}_e}}
\newcommand{\tT}{{\tilde{T}}}
\newcommand{\add}{{\cal A}}
\newcommand{\rem}{{\cal R}}
\newtheorem{sigmodel}{Signal Model}

\newcommand{\thr}{{\text{thr}}}
\newcommand{\delthr}{{\text{del-thr}}}
\newcommand{\delbound}{{b}}
\newcommand{\err}{{\text{err}}}
\newcommand{\Q}{{\cal Q}}

\newcommand{\dett}{{\text{add}}}  
\newcommand{\del}{{\text{del}}}
\newcommand{\CSres}{{\text{CSres}}}
\newcommand{\diff}{{\text{diff}}}
\newcommand{\Section}[1]{ \vspace{-0.13in}  \section{#1} \vspace{-0.12in} } 
\newcommand{\Subsection}[1]{  \vspace{-0.12in} \subsection{#1}  \vspace{-0.08in} } 
\newcommand{\Subsubsection}[1]{   \subsubsection{#1} } 

\date{}
\maketitle

\newcommand{\Aset}{{\cal A}}
\newcommand{\Rset}{{\cal R}}
\newcommand{\Iset}{{\cal I}}
\newcommand{\Dset}{{\cal D}}
\newcommand{\Sset}{{\cal S}}
\newcommand{\Inc}{\text{Inc}}
\newcommand{\Dec}{\text{Dec}}
\newcommand{\Con}{\text{Con}}
\newcommand{\sm}{e}  



\begin{abstract}
In this work, we obtain sufficient conditions for the ``stability" of our recently proposed algorithms, Least Squares Compressive Sensing residual (LS-CS) and modified-CS, for recursively reconstructing sparse signal sequences from noisy measurements. By ``stability" we mean that the number of misses from the current support estimate and the number of extras in it remain bounded by a time-invariant value at all times. We show that, for a signal model with fixed signal power and support set size; support set changes allowed at every time; and gradual coefficient magnitude increase/decrease,  ``stability" holds under mild assumptions -- bounded noise, high enough minimum nonzero coefficient magnitude increase rate, and large enough number of measurements at every time. A direct corollary is that the reconstruction error is also bounded by a time-invariant value at all times. If the support of the sparse signal sequence changes slowly over time, our results hold under weaker assumptions than what simple compressive sensing (CS) needs for the same error bound. Also, our support error bounds are small compared to the support size. Our discussion is backed up by Monte Carlo simulation based comparisons.
\end{abstract}

\section{Introduction} 
The static sparse reconstruction problem has been studied for a while \cite{mallat,bpdn,bdrao_focuss}. The recent papers on compressive sensing (CS)  \cite{donoho,candes,donoho_large,decodinglp,candesnoise,candes_rip} (and many other more recent works) provide the missing theoretical guarantees -- conditions for exact recovery and error bounds when exact recovery is not possible. But for recovering a time sequence of sparse signals, with time-varying sparsity patterns, most existing approaches are batch methods, e.g.  \cite{sparsedynamicMRI,singlepixelvideo}. Our recent work on Least Squares CS-residual (LS-CS) and Kalman filtered CS-residual (KF-CS) \cite{kfcsicip,just_lscs}, and later on modified-CS \cite{isitmodcs0,isitmodcs}, first studied the problem of {\em recursively} recovering a time sequence of sparse signals, with {\em time-varying sparsity patterns}, using {\em much fewer measurements} than what simple CS (CS done at each time separately) needs. By ``recursive" reconstruction, we mean that we want to use only the current measurements' vector and the previous reconstructed signal to reconstruct the current signal. The storage and computational complexity of these solutions is only as much as that of simple CS, but their reconstruction performance is significantly better.
 LS-CS and modified-CS only use the assumption that the sparsity pattern (support in the sparsity basis) changes slowly over time. 
As we show in Fig. \ref{suppchange} and in \cite{isitmodcs}, this is a valid assumption for many medical image sequences. KF-CS also uses slow signal value change.

Denote the support estimate from the previous time by $T$. Modified-CS tries to find a signal that is sparsest outside of $T$ among all signals that satisfy the data constraint. It was first introduced in \cite{isitmodcs0,isitmodcs}, where we studied the noise-free case and obtained exact recovery conditions for it. LS-CS uses a different approach. It replaces CS on the observation by CS on the least squares (LS) residual computed by assuming that $T$ is the correct support \cite{kfcsicip,just_lscs}.
In this work, we obtain the conditions required for ``stability" of LS-CS, modified-CS and of an improved version of modified-CS which we call ``modified-CS with add-LS-del" (improves the support estimation step of modified-CS). By ``stability" we mean that the number of misses from the current support estimate and the number of extras in it remain bounded by a {\em time-invariant} value at all times. A direct corollary is that the reconstruction errors are also bounded by a time-invariant value at all times. 



\subsection{Related Work}

LS-CS and modified-CS are causal and recursive approaches that only rely on the slow support change assumption. Another causal and recursive approach, that uses approximate belief propagation, has been proposed in very recent work \cite{schniter_track}. This is a fully Bayesian approach that assumes prior probabilistic models on both slow support and slow signal value change. Some very interesting numerical experiments are shown.%

``Recursive sparse reconstruction" also sometimes refers to homotopy methods, e.g. \cite{romberg,romberg_jp}, whose goal is to use the past reconstructions and homotopy to speed up the current optimization, but not to achieve accurate recovery from fewer measurements (than what simple CS needs).
Algorithms that improve the reconstruction of a {\em single} signal recursively as more measurements come in, such as those in \cite{seqcs_wilsky,giannakis_jp,romberg_jp}, are also sometimes referred to as ``recursive sparse recovery" algorithms. 
Clearly, the goals in the above works are quite different from ours. 

Also, causal but {\em batch} algorithms for recovering sparse signal sequences, with {\em time-invariant} support,  from fewer measurements were proposed in \cite{giannakis_2}. 

 Other related ideas in literature include the following. Two approaches related to modified-CS are \cite{camsap07} and weighted $\ell_1$ \cite{hassibi}. But both of these focus only on static sparse recovery with prior support knowledge. The work of \cite{hassibi} obtains exact recovery thresholds for weighted $\ell_1$, similar to those in \cite{donoho_large}, for the case when a probabilistic prior on the signal support is available.
 Iterative support estimation approaches (using the recovered support from the first iteration for a second weighted $\ell_1$ step and doing this iteratively) have been studied in recent work \cite{reweighted_cs,iterative_support,hassibi_twostep}. This is done for iteratively improving the recovery of a {\em single} signal.

\begin{figure}
\centerline{
\begin{tabular}{cc}
\epsfig{file = 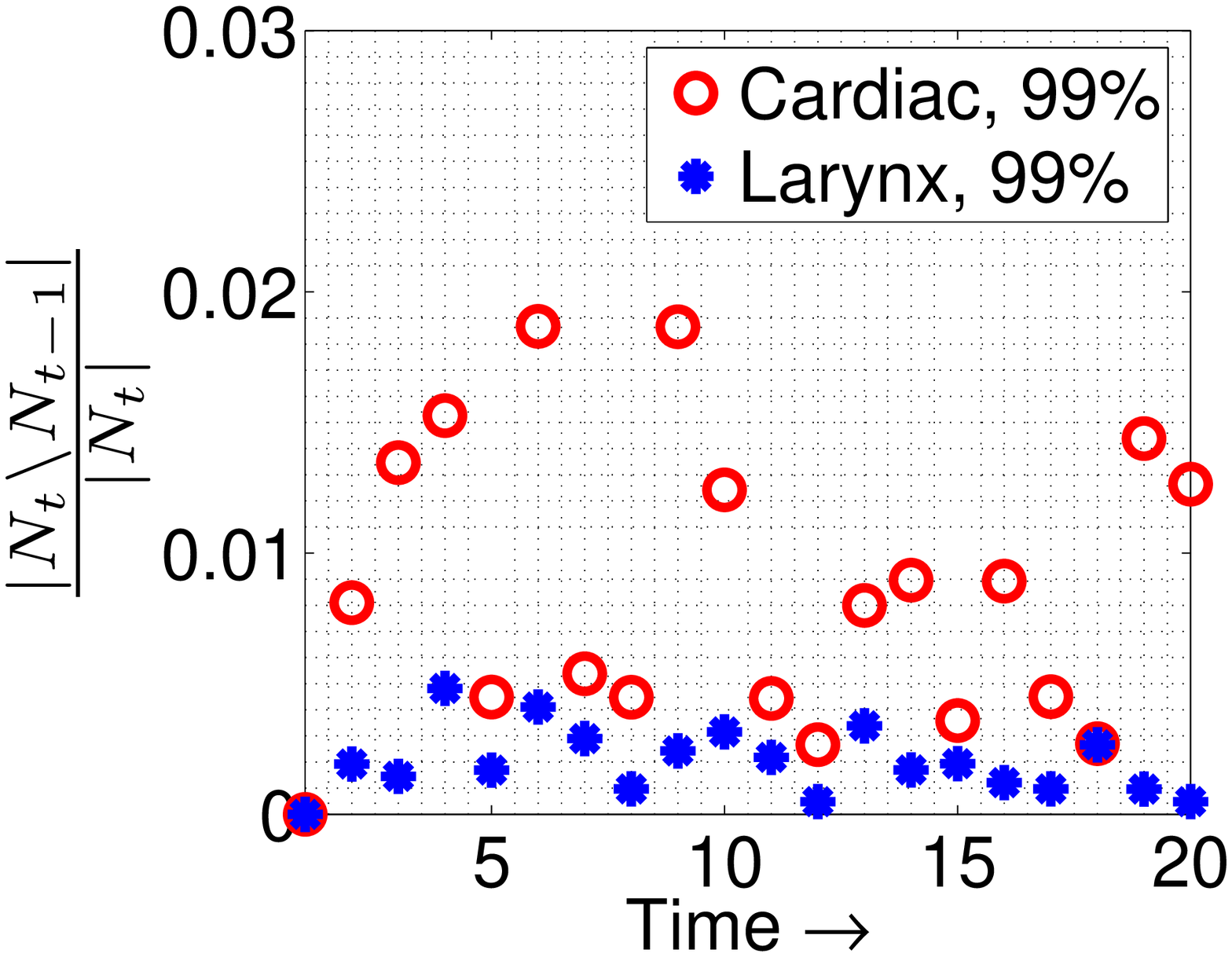,height=2.5cm, width = 4cm} & 
\epsfig{file = 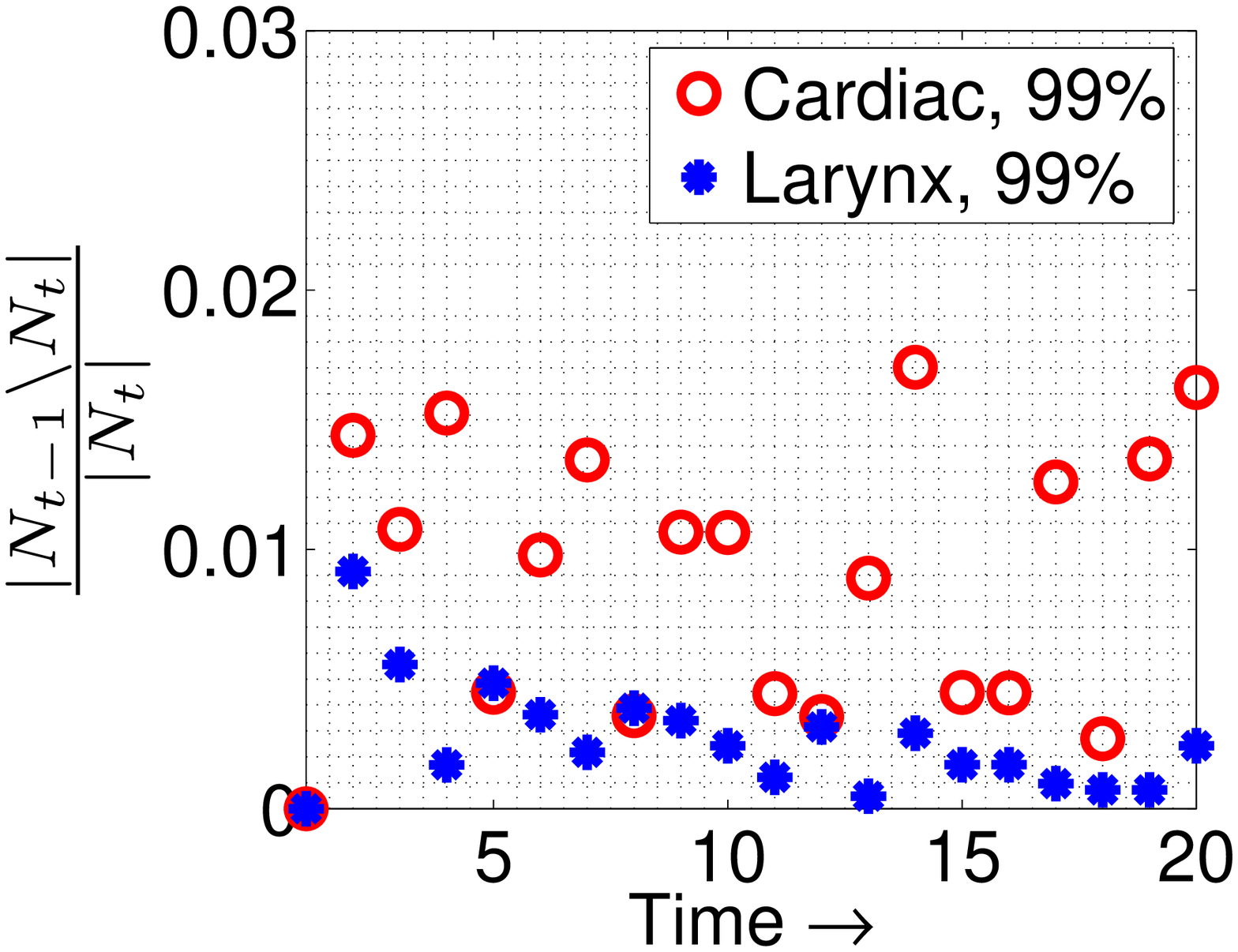,height=2.5cm, width = 4cm}
\end{tabular}
}
\vspace{-0.1in}
\caption{\small{ 
{\bf Slow support change in medical image sequences. }
The two-level Daubechies-4 2D discrete wavelet transform (DWT) served as the sparsity basis.
Since real image sequences are only approximately sparse, we use $N_t$ to denote the 99\%-energy support of the DWT of these sequences. The support size, $|N_t|$, was 6-7\% of the image size for both sequences. 
We plot the number of additions (left) and the number of removals (right) as a fraction of $|N_t|$. {\em Notice that all changes are less than 2\% of the support size.}
}}
\vspace{-0.1in}
\label{suppchange}
\end{figure}

To the best of our knowledge, stability over time has not been studied in the above works for recursive sparse recovery, except in \cite{kfcspap} (KF-CS and LS-CS) or \cite{just_lscs} (LS-CS).
Our result from \cite{kfcspap} is under strong assumptions, e.g. it is for a random walk signal change model (which has unbounded signal power and hence is the easier but unrealistic case), and it requires strong assumptions on the measurement matrix. Our result for LS-CS stability from \cite{just_lscs} holds under mild assumptions and is for a fairly realistic signal change model. The only limitation is that it assumes that support changes occur ``every-so-often" (every $d$ time units, there are $S_a$ support additions and removals). But from testing the slow support change assumption for real data (medical image sequences), it has been observed that support changes usually occur at {\em every} time, e.g. see Fig. \ref{suppchange}. {\em This important case is the focus of the current work.} Moreover, in \cite{just_lscs}, we only studied LS-CS. In this work we study both LS-CS and modified-CS and also modified-CS with add-LS-del.


\subsection{Paper Organization}

%
The paper is organized as follows. We give the problem definition in Sec. \ref{notn} and we overview our results in Sec. \ref{overview}. We describe the signal model that we assume for proving stability in Sec. \ref{signalmodel}. In Sec. \ref{simple_modcs}, we obtain sufficient conditions for the stability of modified-CS and discuss the implications of the result as well as its limitations. 
In Sec. \ref{addLSdel_modcs}, we introduce modified-CS with add-LS-del to address some of the limitations of modified-CS and obtain its stability result. The stability result for modified-CS with add-LS-del is more difficult to obtain because of its improved support estimation procedure. But, in the end the result is also stronger. The result for LS-CS stability is obtained in Sec. \ref{addLSdel_lscs} and compared with previous results. Simulation experiments are discussed in Sec. \ref{sims}. Conclusions are given in Sec. \ref{conclusions}.
The results' overview of Sec. \ref{overview} and some discussions in the later sections can be shortened after review if needed, to make the paper compact.%




\section{Notation, Problem Definition and Overview of Results}
We define notation and give the problem formulation in Sec. \ref{notn}. We give a brief overview of our results in Sec. \ref{overview}.

\subsection{Notation and Problem Definition}
\label{notn}
We let $[1,m]:=[1,2,\dots m]$. We use $T^c$ to denote the complement of a set $T$ w.r.t. $[1,m]$, i.e. $T^c := \{i \in [1,m]: i \notin T \}$. We use $|T|$ to denote the cardinality of $T$. Also, $\emptyset$ denotes the empty set. The set operations $\cup$, $\cap$, $\setminus$ have their usual meanings (recall that $A \setminus B : = A \cap B^c$).

For a vector, $v$, and a set, $T$, $v_T$ denotes the $|T|$ length sub-vector containing the elements of $v$ corresponding to the indices in the set $T$. $\| v \|_k$ denotes the $\ell_k$ norm of a vector $v$. {\em If just $\|v\|$ is used, it refers to $\|v\|_2$.} Similarly, for a matrix $M$, $\|M\|_k$ denotes its induced $k$-norm, while just $\|M\|$ refers to $\|M\|_2$. $M'$ denotes the transpose of $M$ and $M^\dag$ denotes the Moore-Penrose pseudo-inverse of $M$ (when $M$ is tall, $M^\dag:=(M'M)^{-1} M'$). Also, $M_T$ denotes the sub-matrix obtained by extracting the columns of $M$ corresponding to indices in $T$.


At all times, $t>0$, we assume the following observation model:
\bea
y_t = A x_t + w_t,  \ \|w_t\| \le \eps   
\label{obsmod}
\eea
where $x_t$ is an $m$ length sparse vector with support $N_t$; $y_t$ is the $n< m$ length observation vector at time $t$; and $w_t$ is the observation noise. 
As we explain later, our algorithms need more measurements at the initial time, $t=0$. We use $n_0$ to denote the number of measurements used at $t=0$ and we use $A_0$ to denote the corresponding $n_0 \times m$ measurement matrix, i.e. at $t=0$, we have
\bea
y_0 = A_0 x_0 + w_0, \  \|w_0\| \le \eps
\eea

The term ``support", as usual, refers to the set of indices of the nonzero elements of $x_t$.

Our goal is to recursively estimate $x_t$ using $y_1, \dots y_t$. By {\em recursively}, we mean, use only $y_t$ and the estimate from $t-1$, $\xhat_{t-1}$, to compute the estimate at $t$.

The $S$-restricted isometry constant (RIC) \cite{decodinglp}, $\delta_S$, for the matrix, $A$, is the smallest real number satisfying
\bea
(1- \delta_S) \|c\|^2 \le \|A_T c\|^2 \le (1 + \delta_S) \|c\|^2
\label{def_delta}
\eea
for all sets $T \subset [1,m]$ of cardinality $|T| \le S$ and all real vectors $c$ of length $|T|$.
The restricted orthogonality constant (ROC) \cite{decodinglp}, $\theta_{S_1,S_2}$, is the smallest real number satisfying
\bea
| {c_1}'{A_{T_1}}'A_{T_2} c_2 | \le \theta_{S_1,S_2} \|c_1\| \ \|c_2\|
\label{def_theta}
\eea
for all disjoint sets $T_1, T_2 \subset [1,m]$ with $|T_1| \le S_1$, $|T_2| \le S_2$ and $S_1+S_2 \le m$, and for all vectors $c_1$, $c_2$ of length $|T_1|$, $|T_2|$ respectively.

{\em In this work, $\delta_{S}$, $\theta_{S_1,S_2}$ always refer to the RIC, ROC for the measurement matrix $A$ which is used at $t>0$. If we refer to the RIC of any other matrix, e.g. $A_0$, we use $\delta_S(A_0)$.}


We use $\alpha$ to denote the support estimation threshold used by modified-CS and we use $\alpha_{\dett}, \alpha_{\del}$ to denote the support addition and deletion thresholds used by modified-CS with add-LS-del and by LS-CS.
We use $\Nhat_t$ to denote the support estimate at time $t$. To keep notation simple, we avoid using the subscript $t$ wherever possible.

\bd[$T_t$, $\Delta_t$, $\Delta_{e,t}$]
We use $T_t  : = \Nhat_{t-1}$ to denote the support estimate from the previous time. This serves as the predicted support at time $t$.
We use $\Delta_t := N_t \setminus T_t$ to denote the unknown part of $T_t$ and $\Delta_{e,t} := T_t \setminus N_t$ to denote the ``erroneous" part of $T_t$. In many places in the manuscript, we remove the subscript $t$ to keep notation simple.
\ed
With the above definition, clearly,
$$N_t = T_t \cup \Delta_t \setminus \Delta_{e,t}.$$

\bd[$\tT_t$, $\tDelta_t$, $\tDelta_{e,t}$]
We use $\tT_t := \Nhat_t$ to denote the final estimate of the current support; $\tDelta_t : = N_t \setminus \tT_t$ to denote the ``misses" in $\Nhat_t$ and $\tDelta_{e,t} := \tT_t \setminus N_t$ to denote the ``extras". 
\ed

We sometimes refer to $\Delta, \Delta_e$ as the predicted support errors and to $\tDelta, \tDelta_e$ as the final (or estimated) support errors.
The sets $T_{\dett}, \Delta_{\dett}, \Delta_{e,\dett}$ are defined in Definition \ref{defdett} (Sec. \ref{addLSdel_modcs}).%

{\em If two sets $B$, $C$ are disjoint, we just write $D \cup B \setminus C$ instead of writing $(D \cup B) \setminus C$, e.g. $N_t = T \cup \Delta \setminus \Delta_e$.}

We refer to the left (right) hand side of an equation or inequality as LHS (RHS).

In this work, ``modified-CS" refers to the solution of (\ref{modcs}). Also, simple CS refers to  the solution of (\ref{modcs}) with $T=\emptyset$.

%

\vspace{-0.1in}
\subsection{Overview of Results}
\label{overview}

When measurements are noisy, the reconstruction errors of modified-CS and of LS-CS can easily be bounded as a function of the support size, $|N_t|$, and of the predicted support error sizes, $|\Delta_t|$ and $|\Delta_{e,t}|$ \cite{arxiv,just_lscs}.
The bound is small at time $t$ if $|\Delta_t|$ and  $|\Delta_{e,t}|$ are small enough. But smallness of the predicted support errors, $\Delta_t$, $\Delta_{e,t}$, depends on the accuracy of the previous reconstruction, and thus, in general, it may happen that, over time, the error bound keeps increasing. Such a result is of limited use for a recursive reconstruction problem. There is thus a need to obtain conditions under which one can show ``stability", i.e. ensure that a time-invariant bound holds on the sizes of these support errors. Combining this with the error bound result will imply that the reconstruction error is also bounded by a time-invariant value at all times.



In this work, we obtain results for the stability of three algorithms: (a) modified-CS; (b) ``modified-CS with add-LS-del" and (c) LS-CS. ``Modified-CS with add-LS-del" improves the support estimation step of modified-CS by using a three step approach first introduced in \cite{kfcsicip,just_lscs} and in \cite{subspacepursuit,cosamp} -- support addition with a smaller threshold, followed by LS estimation on the new support, and finally support deletion using the LS estimate. Using add-LS-del significantly improves both the stability result we can prove (as argued in Sec. \ref{modcsald_discuss}) and the empirical reconstruction performance we get (see Sec. \ref{sims}). 

All our results are obtained under a bounded observation noise assumption and for a signal model with
\ben
\item support changes ($S_a$ additions and $S_a$ removals) occurring at every time, $t$,
\item magnitude of the newly added coefficients increasing gradually, and similarly for decrease before removal,
\item support size, $|N_t|=S_0$ at all times and the signal power\footnote{Usually signal power refers to the expected value of the 2-norm of the signal, $\E[\|x_t\|^2]$. In our work, we assume a deterministic signal model and hence signal power just refers to $\|x_t\|^2$.}, $\|x_t\|^2$, also constant at all times.
\een

Our results have the following form. For a given number and type of measurements (i.e. for a given measurement matrix, $A$), and for a given noise bound, $\eps$, if,
\ben
\item the support estimation threshold(s) is/are appropriately set,
\item the support size, $S_0$, and the newly added (or removed) support size, $S_a$, are small enough,
\item the newly added coefficients' increase rate (existing large coefficients' decrease rate), $r$, is large enough,  and
\item the initial number of measurements, $n_0$, is large enough for accurate initial reconstruction using simple CS,
\een
then, the support error sizes are bounded by time-invariant values: we show that $|\tDelta_t| \le 2S_a$, $|\tDelta_{e,t}| = 0$ and $|\Delta_t| \le 2S_a$, $|\Delta_{e,t}| \le S_a$. A direct corollary is that the reconstruction error is also bounded by a time-invariant value at all times.

\begin{remark}
The reason we need to assume bounded noise is as follows. When the noise is unbounded, e.g. Gaussian, all error bounds for CS and, similarly, all error bounds for LS-CS or modified-CS hold with ``large probability", e.g. see \cite{dantzig,just_lscs}. For stability, we need the error bound for LS-CS or modified-CS to hold at all times, $0 \le t < \infty$ (this, in turn, is used to ensure that the support gets estimated with bounded error at all times). Clearly, this will be a zero probability event.
\\
As an aside, most existing works which use the RIC based approach of Candes et al to bound the error of noisy sparse recovery, or of noisy sparse recovery with partial support knowledge, also assume bounded noise, e.g. \cite{candesnoise,candes_rip,arxiv}.%
\label{boundednoise}
\end{remark}

\begin{remark}
We should mention that constant or bounded signal power is both the more practical case (since, in practice, signal power never keeps increasing unboundedly) and is also the more difficult case. This is because the accuracy of the reconstruction at time $t+1$ relies heavily on the correct detection of the small elements at time $t$. Correct detection will become easier for larger signal power (or, to be precise, for larger  power of the smallest nonzero coefficients).
\label{sigpow}
\end{remark}

For our signal model, slow support change translates to $S_a \ll S_0$. Under this assumption, clearly, $2S_a \ll S_0$, and so our support error bounds are small compared to the support size, $S_0$, making our stability results meaningful. We can argue that our results hold under weaker assumptions (allow larger values of $S_0$), for a given measurement matrix $A$, than the corresponding simple CS (CS done at each time separately) result. Since simple CS is not a recursive approach, the CS error bound from \cite{candes_rip} (or other works) also serves as a stability result for it. Also, we can argue that modified-CS with add-LS-del needs the weakest conditions on the number of measurements, $n$, and on the rate of coefficient magnitude increase/decrease, $r$. Modified-CS needs similar conditions on $n$, but needs a larger $r$. LS-CS needs the strongest conditions on both $n$ and $r$. Since we can only compare sufficient conditions or upper bounds, we back up all our discussion with simulation experiments to compare actual reconstruction performance.%


\section{Signal Model for Studying Stability}
\label{signalmodel}
The modified-CS or LS-CS algorithms {\em do not} assume any signal model. But for showing stability, we need certain assumptions on the signal change over time.

\begin{sigmodel} Assume the following.
\ben

\item {\em (addition) } At each $t>0$, $S_a$ new coefficients get added to the support at magnitude $r$. Denote this set by $\Aset_t$.

\item {\em (increase) } At each $t>0$, the magnitude of $S_a$ coefficients out of all those which had magnitude $(j-1)r$ at $t-1$ increases to $jr$. This occurs for all $2 \le j \le d$. Thus the maximum magnitude reached by any coefficient is $M:=dr$.

\item {\em (decrease) } At each $t>0$, the magnitude of $S_a$ coefficients out of all those which had magnitude $(j+1)r$ at $t-1$ decreases to $jr$. This occurs for all $1 \le j \le (d-2)$.

\item {\em (removal) } At each $t>0$, $S_a$ coefficients out of all those which had magnitude $r$ at $t-1$ get removed from the support (magnitude becomes zero). Denote this set by $\Rset_t$.

\item {\em (initial time) } At $t=0$, the support size is $S_0$ and it contains $2S_a$ elements each with magnitude $r,2r, \dots (d-1)r$, and $(S_0-(2d-2)S_a)$ elements with magnitude $M$.%
\een
\label{sigmod2}
\end{sigmodel}

\begin{figure}
\centerline{
\epsfig{file = 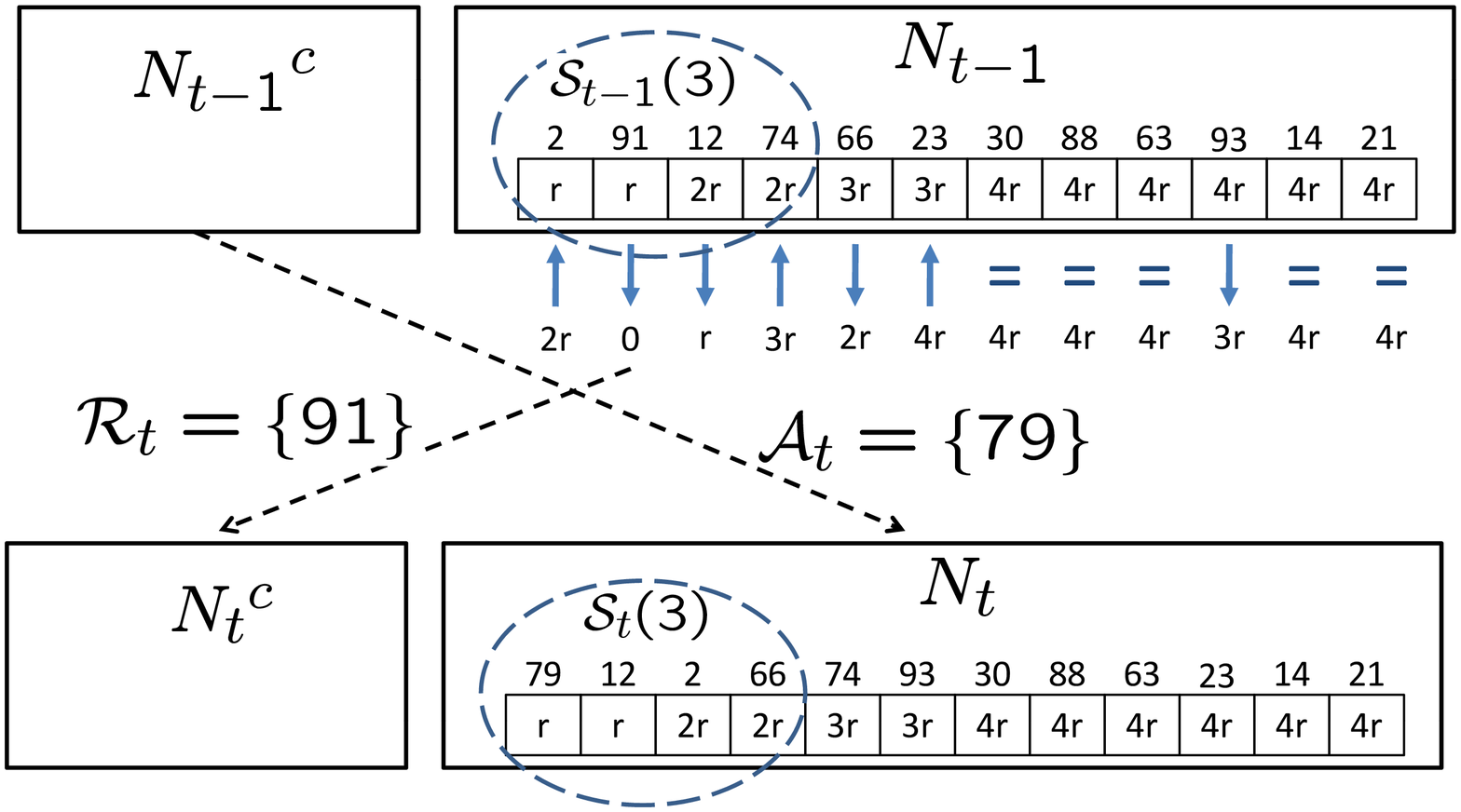, width=9cm, height=4cm}
}
\vspace{-0.1in}
\caption{\small{
An example of Signal Model \ref{sigmod2} with $m=100$, $S_0=12$, $S_a=1$, and $d=4$. Thus at any time it contains $2S_a=2$ elements each with magnitude $r$, $2r$, and $3r$ and $S_0-(2d-2)S_a=6$ elements with stable magnitude $M=4r$. We show each support element's magnitude inside a square box and its index just above the box. The up and down arrows below the $N_{t-1}$ box indicate whether the element increases or decreases. An ``=" indicates that the element magnitude remains constant at $4r$. In both $N_{t-1}$ and $N_t$ we have circled the small elements' set $\Sset_{t-1}(3)$ and $\Sset_{t}(3)$ respectively.
}}
\vspace{-0.1in}
\label{signalmodelfig}
\end{figure}

We show an example of the above signal model in Fig. \ref{signalmodelfig}.

The above model has the following realistic features --  (a) equal number, $S_a$, of additions and removals to (from) the support occur at every time, $t$; (b) a newly added coefficient gets added at a small magnitude; (c) magnitude of any nonzero element either remains constant, or increases gradually at rate $r$, but not beyond a maximum magnitude $M:=dr$, or decreases gradually at rate $r$; and (d) at all times, the signals have the same support set size, $|N_t|=S_0$ and the same signal power, $\|x_t\|^2=(S_0-(2d-2)S_a)M^2 + 2S_a\sum_{j=1}^{d-1} j^2 r^2$ 
%


In practice, the number of additions/removals to the support is never exactly equal, but varies in a small range over time. A similar thing holds for the coefficient increase/decrease rate, $r$, or for the stable magnitude, $M$. But for notational simplicity, we ignore these variations \footnote{To model the  variations over time compactly, a probabilistic signal change model will be a better one. But that will make our analysis a lot more tricky since the reconstruction error bounds, which form the starting point for our stability results, do not assume any randomness \cite{candes_rip,just_lscs,arxiv}. In particular, they do not treat the sparse signal as a random variable.}. Also, in practice, different nonzero elements may have different magnitude increase rates, $r_i$, and different stable magnitudes, $M_i$. It will be possible to extend our results to this latter case fairly easily, and we expect that the result will require a lower bound on $\min_i r_i$. 


Signal Model \ref{sigmod2} does not specify a particular generative model. Two examples of signal models that satisfy the above assumptions are given in Appendix \ref{generativemodel}. Briefly, in the first model, at each $t$, $S_a$ new elements, randomly selected from ${N_{t-1}}^c$, get added to the support at initial magnitude, $r$, and equally likely sign. Their magnitude keeps increasing gradually, at rate $r$, for  $d$ time units after which it becomes constant at $M:=dr$. The sign does not change. Also, at each time, $t$, $S_a$ randomly selected elements out of the ``stable" elements' set (set of elements which have magnitude $M$ at $t-1$), begin to decrease at rate $r$ and this continues until their magnitude becomes zero, i.e. they get removed from the support.  
A second possible generative model randomly selects $S_a$ out of the $2S_a$ current elements with magnitude $jr$ and increases them, and decreases the other $S_a$ elements. 

To understand the implications of the assumptions in Signal Model \ref{sigmod2}, we define the following sets.
\bd
Define the following.
\ben
\item For all $0 \le j \le d-1$, let $$\Dset_{t}(j): = \{i: |x_{t,i}| = jr, \ |x_{t-1,i}| = (j+1)r \}$$
denote the set of elements that {\em decrease} from $(j+1)r$ to $jr$ at time, $t$.
\item For all $1 \le j \le d$, let $$\Iset_{t}(j): = \{i: |x_{t,i}| = jr, \ |x_{t-1,i}| = (j-1)r \}$$ denote the set of elements that {\em increase} from $(j-1)r$ to $jr$ at time, $t$.
\item For all $1 \le j \le d-1$, let  $$\Sset_t(j):= \{i:  0 < |x_{t,i}| < j r \}$$ denote the set of {\em small but nonzero} elements, with smallness threshold $jr$.
\item Clearly,
\ben
\item The newly added set, $$\Aset_t= \Iset_t(1)$$
\item The newly removed set, $$\Rset_t= \Dset_t(0)$$
\item $|\Iset_{t}(j)|=S_a$, $|\Dset_{t}(j)|=S_a$, $|\Sset_t(j)| = 2(j-1)S_a$.
\een
\een
\label{defIset}
\ed

Consider a $1 < j \le d$. From Signal Model \ref{sigmod2}, it is clear that at any time, $t$, $S_a$ elements enter the small elements' set, $\Sset_t(j)$, from the bottom (set $\Aset_t$) and $S_a$ enter from the top (set $\Dset_{t}(j-1)$). Similarly $S_a$ elements leave  $\Sset_t(j)$ from the bottom  (set $\Rset_t$) and $S_a$ from the top (set $\Iset_{t}(j)$). Thus,
\bea
\Sset_t(j) = \Sset_{t-1}(j)  \cup (\Aset_t \cup \Dset_{t}(j-1)) \setminus (\Rset_t \cup \Iset_{t}(j)) \ \ \
\label{sseteq}
\eea
To look at an example, see Fig. \ref{signalmodelfig}. Consider $j=3$. Notice that $\Sset_{t-1}(3) =  \{2,91,12,74\}$ and $\Sset_t(3) = \{79,12,2,66\}$. Also, $\Aset_t = \{79\}$, $\Rset_t = \{91\}$, $\Iset_t(3) = \{74\}$ and $\Dset_t(2) = \{66\}$. Clearly $\{2,91,12,74\} \cup (\{79\} \cup  \{66\}) \setminus (\{91\} \cup  \{74\} = \{79,12,2,66\}$, i.e. (\ref{sseteq}) holds.

Since $\Aset_t, \Rset_t, \Dset_{t}(j-1),\Iset_{t}(j)$ are mutually disjoint, $\Rset_t \subseteq \Sset_{t-1}(j)$ and $\Iset_{t}(j) \subseteq \Sset_{t-1}(j)$, thus, (\ref{sseteq}) implies that
\bea
\Sset_{t-1}(j)  \cup \Aset_t  \setminus \Rset_t = \Sset_t(j) \cup \Iset_{t}(j) \setminus \Dset_{t}(j-1)
\label{sseteq_2}
\eea

Also, clearly, from Signal Model \ref{sigmod2},
\bea
N_t \se N_{t-1} \cup \Aset_t \setminus \Rset_t
\eea
We will use these in the proof of the results of Sec. \ref{addLSdel_modcs}.

\section{Stability of modified-CS}
\label{simple_modcs}
Modified-CS was first proposed in \cite{isitmodcs0,isitmodcs} as a solution to the problem of sparse reconstruction with partial, and possibly erroneous, knowledge of the support. Denote this ``known" support by $T$. Modified-CS tries to find a signal that is sparsest outside of the set $T$ among all signals satisfying the data constraint. In the noisy case, it solves $\min_\beta  \|(\beta)_{T^c}\|_1 \ \text{s.t.} \ \| y_t - A \beta \| \le \eps$. For recursively reconstructing a time sequence of sparse signals, we use the support estimate from the previous time, $\Nhat_{t-1}$, as the set $T$. The support is estimated by thresholding the output of modified-CS. At the initial time, $t=0$, we let $T$ be the empty set, $\emptyset$, i.e. we do simple CS. Alternatively, as explained in \cite{isitmodcs}, we can use prior knowledge of the initial signal's support as the set $T$ at $t=0$, e.g. for wavelet sparse images with no (or a small) black background, the set of indices of the approximation coefficients can form the set $T$. This prior knowledge is usually not as accurate. Thus, in either case, at $t=0$ we need more measurements, i.e. $n_0 > n$.

In this work, for simplicity, we assume that simple CS is done at $t=0$. We summarize the algorithm in Algorithm \ref{modcsalgo}.




\begin{algorithm}[h!]
\caption{{\bf \small Modified-CS}}
For $t \ge 0$, do
\ben
\item {\em Simple CS. } If $t = 0$, set $T = \emptyset$ and compute $\xhat_{t,modcs}$ as the solution of
\bea
\min_\beta  \|(\beta)\|_1 \ \text{s.t.} \ \| y_0 - A_0 \beta \| \le \eps
\label{simpcs}
\eea

\item {\em Modified-CS. } If $t>0$, set $T = \Nhat_{t-1}$ and compute $\xhat_{t,modcs}$ as the solution of
\label{step1_0}
\bea
\min_\beta  \|(\beta)_{T^c}\|_1 \ \text{s.t.} \ \| y_t - A \beta \| \le \eps
\label{modcs}
\eea


\item {\em Estimate the Support. } Compute $\tT$ as
\label{supp_estim}
\bea
\tT=\{i \in [1,m] : |(\xhat_{t,modcs})_i| > \alpha \}
\eea

\item Set $\Nhat_t = \tT$. Output $\hat{x}_{t,modcs}$. Feedback $\Nhat_t$.
\een
\label{modcsalgo}
\end{algorithm}

By adapting the approach of \cite{candes_rip}, the error of modified-CS can be bounded as a function of $|T|=|N|+|\Delta_e|-|\Delta|$ and $|\Delta|$. This was done in \cite{arxiv}. We state a modified version here.

\begin{lemma}[modified-CS error bound]
Let $x$ be a sparse vector with support $N$ and let $y:=Ax+w$ with $\|w\| \le \eps$. Also, let $\Delta:=N \setminus T$ and $\Delta_e:=T \setminus N$. Let $\xhat_{modcs}$ denote the solution of (\ref{modcs}).
If
\bi
\item $\delta_{|N|+|\Delta|+|\Delta_e|} < \sqrt{2}-1$ and $|\Delta| \le |N|/3$,
\ei
then
\bea
\|x - \xhat_{modcs}\| \sle C_1(|N|+|\Delta|+|\Delta_e|) \eps, \ \text{where} \nn \\
C_1(S) \sdefn \frac{4 \sqrt{1+\delta_S}}{1 - (\sqrt{2} +1) \delta_S}
\label{defC1s}
\eea
\label{modcsbnd}
\end{lemma}
For the sake of completeness, and for ease of review, we provide a proof in the last appendix, Appendix \ref{modcserrorbound}. This can later be removed.%

If $\delta_{|N|+|\Delta|+|\Delta_e|}$ is just smaller than $\sqrt{2}-1$, the error bound will be very large because the denominator of $C_1(S)$ will be very large. To keep the bound small, we need to assume that $\delta_{|N|+|\Delta|+|\Delta_e|} < b(\sqrt{2}-1)$ with a $b<1$. For simplicity, let $b=1/2$. Then we get the following corollary, which we will use in our stability results.

\begin{corollary}[modified-CS error bound]
Let $x$ be a sparse vector with support $N$ and let $y:=Ax+w$ with $\|w\| \le \eps$. Also, let $\Delta:=N \setminus T$ and $\Delta_e:=T \setminus N$.
Let $\xhat_{modcs}$ denote the solution of (\ref{modcs}). If
\label{modcs_cs_bnd}
\bi
\item $\delta_{|N|+|\Delta|+|\Delta_e|} < (\sqrt{2}-1)/2$ and $|\Delta| \le |N|/3$,
\ei
then
\bea
\|x - \xhat_{modcs}\| \sle C_1(|N|+|\Delta|+|\Delta_e|) \eps \le 8.79 \eps
\eea
\end{corollary}

{\em Proof: } Notice that $C_1(S)$ is an increasing function of $\delta_S$. The above corollary follows by using $\delta_{|N|+|\Delta|+|\Delta_e|} < (\sqrt{2}-1)/2$ to bound $C_1(S)$ by $C_1((\sqrt{2}-1)/2)=8.79$.
%

We can state a similar version of the result for CS \cite{candes_rip}. 

\begin{corollary}[CS error bound \cite{candes_rip}]
Let $x$ be a sparse vector with support $N$ and let $y:=Ax+w$ with $\|w\| \le \eps$. Let $\xhat_{cs}$ denote the solution of (\ref{modcs}) with $T = \emptyset$. If 
\label{cs_bnd}
\bi
\item $\delta_{2|N|} < (\sqrt{2}-1)/2$,
\ei
then
\bea
\|x - \xhat_{cs}\| \sle C_1(2|N|) \eps \le 8.79 \eps
\eea
\end{corollary}

\subsection{Stability result for modified-CS}
\label{stab_modcs}

The first step to show stability is to find sufficient conditions for a certain set of large coefficients to definitely get detected, and for the elements of $\Delta_e$ to definitely get deleted. These can be obtained using Corollary  \ref{modcs_cs_bnd} and the following simple facts which we state as a proposition.

\begin{proposition}[simple facts]Consider Algorithm \ref{modcsalgo}.%
\ben
\item An $i \in N$ will definitely get detected in step \ref{supp_estim} if $|x_i|  > \alpha + \|x - \xhat_{modcs}\|$. This follows since  $ \|x - \xhat_{modcs}\| \ge \|x - \xhat_{modcs}\|_\infty \ge |(x - \xhat_{modcs})_{i}|$.
\label{det1}


\item Similarly, all $i \in \Delta_{e}$ (the zero elements of $T$) will definitely get deleted in step \ref{supp_estim} if $\alpha \ge \|x - \xhat_{modcs}\|$.

\een
\label{prop0}
\end{proposition}

Combining the above facts with Corollary \ref{modcs_cs_bnd}, we get the following lemma.

\begin{lemma} 
Let $x$ be a sparse vector with support $N$ and let $y:=Ax+w$ with $\|w\| \le \eps$. Also, let $\Delta:=N \setminus T$ and $\Delta_e:=T \setminus N$.
\\ Assume that $|N| = S_N$, $|\Delta_e| \le S_{\Delta e}$ and $|\Delta| \le S_\Delta$.
\\ Consider Algorithm \ref{modcsalgo}.
\ben
\item Let $L:=\{i \in N: |x_i| \ge b_1 \}$. All elements of $L$ will get detected in step \ref{supp_estim} if
\ben
\item $\delta_{S_N + S_{\Delta e} + S_\Delta} < (\sqrt{2}-1)/2$ and $S_\Delta  \le S_N/3$, and
\item $b_1 > \alpha +  8.79 \eps$.
\een

\item In step \ref{supp_estim}, there will be no false additions, and all the true removals from the support (the set $\Delta_{e}$) will get deleted at the current time, if
\ben
\item $\delta_{S_N + S_{\Delta e} + S_\Delta} < (\sqrt{2}-1)/2$ and $S_\Delta \le S_N/3$, and%
\item $\alpha \ge 8.79 \eps$. 
\een

\een
\label{lemma_modcs}
\end{lemma}

In the above lemma and proposition, for ease of notation, we have removed the subscript $t$ from $x_t$, $N_t$, $T_t$ and $\Delta_t$.

We use the above lemma to obtain the stability result as follows. Let us fix a bound on the maximum allowed magnitude of a missed coefficient.
Suppose we want to ensure that only coefficients with magnitude less than $2r$ are part of the final set of misses, $\tDelta_t$, at any time, $t$ and that the final set of extras, $\tDelta_{e,t}$ is an empty set. In other words, we want to find conditions to ensure that $\tDelta_t \subseteq \Sset_{t}(2)$ and $|\tDelta_{e,t}|=0$. Using Signal Model \ref{sigmod2}, $|\Sset_{t}(2)|=2S_a$ and thus $\tDelta_t \subseteq \Sset_{t}(2)$ will imply that $|\tDelta_t| \le 2S_a$. This leads to the following result. The result can be easily generalized to ensure that, for some $d_0 \le d$, $\Delta_t \subseteq \Sset_t(d_0)$, and thus $|\Delta_t| \le (2d_0-2)S_a$, holds at all times $t$. We show how to do this for the result of the next section in Appendix \ref{stabres_modcs_gen}; an analogous thing can be done for Theorem \ref{stabres_simple_modcs} as well.%

\begin{theorem}[Stability of modified-CS]
Assume Signal Model \ref{sigmod2} on $x_t$. Also assume that $y_t$ satisfies (\ref{obsmod}) with $\|w_t\| \le \eps$. 
Consider Algorithm \ref{modcsalgo}. If the following hold
\ben

\item {\em (support estimation threshold) } set $\alpha =   8.79 \eps$ 
\label{threshes_simple}

\item {\em (support size, support change rate)} $S_0$, $S_a$ satisfy $\delta_{S_0 + 3S_a} < (\sqrt{2}-1)/2$ and $S_a \le S_0/6$,
\label{measmodel_simple}

\item {\em (new element increase rate) } $r \ge G$, where
\label{add_del_simple}
\bea
G \sdefn \frac{ \alpha + 8.79 \eps }{2} = 8.79 \eps
\eea

\item {\em (initial time)} at $t=0$, $n_0$ is large enough to ensure that $\tDelta_0  \subseteq \Sset_0(2)$, $|\tDelta_0| \le 2S_a$,  $|\tDelta_{e,0}| =0$ and $|\tT_0| \le S_0$
\label{initass_simple}
\een
then,
\ben
\item at all $t \ge 0$, $|\tT_t| \le S_0$, $|\tDelta_{e,t}| =0$, $\tDelta_t \subseteq \Sset_t(2)$ and so $|\tDelta_t| \le 2S_a$,
\item at all $t > 0$, $|T_t| \le S_0$, $|\Delta_{e,t}| \le S_a$, and $|\Delta_t| \le 2S_a$,
\item at all $t > 0$,  $\|x_t - \xhat_{t,modcs}\| \le 8.79 \eps$ 
\een
\label{stabres_simple_modcs}
\end{theorem}

{\em Proof: } The complete proof is given in Appendix \ref{proof_simple_modcs}. It follows using induction. We use the induction assumption; the fact that $T_t = \tT_{t-1} = \Nhat_{t-1}$; and the fact that $N_t = N_{t-1} \cup \Aset_t \setminus \Rset_t$ to bound $|T_t|$, $|\Delta_t|$ and $|\Delta_{e,t}|$. Next, we use these bounds and Lemma \ref{lemma_modcs} to bound $|\tDelta_t|$ and $|\tDelta_{e,t}|$. Finally  $|\tT_t| \le |N_t| +  |\tDelta_{e,t}|$ helps to bound $|\tT_t|$.

\subsection{Discussion}
\begin{remark}
We note that condition \ref{initass} is not restrictive. It is easy to see that this will hold if the number of measurements at $t=0$, $n_0$, is large enough to ensure that the measurement matrix at $t=0$, $A_0$, satisfies $\delta_{2S_0}(A_0) < (\sqrt{2}-1)/2$ and conditions \ref{threshes_simple} and \ref{add_del_simple} hold.
\label{initass_remark}
\end{remark}


Notice that all the support errors are bounded by $2S_a$ or less. Under slow support change, $S_a \ll S_0$ and so $2S_a$ is also small compared to the support size, $S_0$, making the above result a {\em meaningful} stability result.

Let us compare the results for modified-CS and simple CS. Since simple CS is not a recursive approach (each time instant is handled separately), Corollary \ref{cs_bnd} is also a stability result for it. From Corollary \ref{cs_bnd}, simple CS needs $\delta_{2S_0} < (\sqrt{2}-1)/2$ to ensure that its error is bounded by $8.79 \eps$ for all $t$. On the other hand, for $t>0$, our result from Theorem \ref{stabres_simple_modcs} only needs $S_a \le S_0/6$ and $\delta_{S_0 + 3S_a} < (\sqrt{2}-1)/2$ to get the same error bound.
Under $S_a \ll S_0$ (slow support change), $S_a \le S_0/6$ easily holds and $\delta_{S_0 + 3S_a} < (\sqrt{2}-1)/2$ is clearly weaker than the simple CS condition. Thus, at $t>0$, for a given measurement matrix $A$, modified-CS error is guaranteed to remain below $8.79\eps$ for larger support sizes, $S_0$, than for simple CS.  Said another way, for a given $S_0$, modified-CS needs fewer measurements (only enough to satisfy $\delta_{S_0 + 3S_a} < (\sqrt{2}-1)/2$), than simple CS (which needs enough to satisfy $\delta_{2S_0} < (\sqrt{2}-1)/2$).

At $t=0$, the modified-CS algorithm of Algorithm \ref{modcsalgo} needs the same number of measurements as simple CS. If reliable prior support knowledge were available at $t=0$, one would need fewer measurements even at $t=0$.

The above discussion only compares sufficient conditions. We back it up with actual simulation comparisons in Fig. \ref{simfig_n65_r1} and \ref{simfig_n59_r1} where we compare the average reconstruction error when $n$ is just large enough to ensure small (less than 0.5\%) error for modified-CS. With this $n$, the CS error is between 20-30\%. Here ``error" refers to normalized mean squared error (NMSE). The simulation details are given in Sec. \ref{sims}.

\subsection{Limitations}
\label{limitations}
Before going further, let us discuss the limitations of the above result and of modified-CS itself. First, in Proposition \ref{prop0}, and hence everywhere after that, we bound the $\ell_\infty$ norm of the error by the $\ell_2$ norm. This is often a loose bound and results in a loose lower bound on the required threshold $\alpha$ and consequently a larger than required lower bound on the minimum required rate of coefficient increase/decrease, $r$.

Second, modified-CS uses single step thresholding for estimating the support $\Nhat_t$. The threshold, $\alpha$,  needs to be large enough to ensure correct deletion of all the removed elements and no false detection of zero elements (condition \ref{threshes_simple}). But this means that the magnitude increase rate, $r$, needs to be even larger to ensure correct detection, and no false deletion, of all but the smallest $2S_a$ nonzero elements (condition \ref{add_del_simple}).


There is another related issue which is not seen in the theoretical analysis because we only bound the $\ell_2$ norm of the error, but is actually more important since it affects the reconstruction itself, not just the sufficient conditions for its stability. This has to do with the fact that $\xhat_{t,modcs}$ is a biased estimate of $x_t$. A similar issue for noisy CS, and a possible solution (Gauss-Dantzig selector), was first discussed in \cite{dantzig}. In our context, along $T^c$, the values of $\xhat_{t,modcs}$ will be biased towards zero (because we minimize $\|(\beta)_{T^c}\|_1$), while, along $T$, they may be biased away from zero (since there is no constraint on $(\beta)_T$). The bias will be larger when the noise is larger. This will create the following problem. The set $T$ contains the set $\Delta_e$ which needs to be deleted. Since the estimates along $\Delta_e$ may be biased away from zero, one will need a higher threshold to delete them. But that would make detection more difficult, especially since the estimates along $\Delta \subseteq T^c$ will be biased towards zero. In the next section, we discuss a partial solution to this and the previous issue.

\section{Modified-CS with Add-LS-Del and its Stability}
\label{addLSdel_modcs}

The last two issues mentioned above in Sec. \ref{limitations} can be partly addressed by replacing the single support estimation step by a three step Add-LS-Del procedure summarized in Algorithm \ref{modcsalgo_2}.  This idea was first introduced in our older work \cite{just_lscs,kfcsicip} for recursive sparse reconstruction and simultaneously also in \cite{subspacepursuit,cosamp} for greedy algorithms for static sparse reconstruction.
It involves a support addition step (that uses a smaller threshold), as in (\ref{addstep}), followed by LS estimation on the new support estimate, $T_\dett$, as in (\ref{lsstep}), and then a deletion step that thresholds the LS estimate, as in (\ref{deletestep}). This can be followed by a second LS estimation using the final support estimate, as in (\ref{finalls}), although this last step is not critical. The addition step threshold, $\alpha_{\dett}$, needs to be just large enough to ensure that the matrix used for LS estimation, $A_{T_\dett}$ is well-conditioned. If $\alpha_{\dett}$ is chosen properly and if $n$ is large enough, the LS estimate on $T_\dett$ will have smaller error than the modified-CS output. As a result, deletion will be more accurate when done using this estimate. This also means that one can also use a larger deletion threshold, $\alpha_{\del}$, which will ensure quicker deletion of extras.
We summarize the algorithm in Algorithm \ref{modcsalgo_2}. Notice the reduction in error of modified-CS with add-LS-del as compared to modified-CS in Fig. \ref{fig1}.


\begin{algorithm}[h]
\caption{{\bf \small Modified-CS with Add-LS-Del}}
For $t \ge 0$, do
\ben
\item {\em Simple CS. } If $t = 0$, set $T = \emptyset$ and compute $\xhat_{t,modcs}$ as the solution of (\ref{simpcs}).

\item {\em Modified-CS. } If $t>0$, set $T = \Nhat_{t-1}$ and compute $\xhat_{t,modcs}$ as the solution of (\ref{modcs}).
\label{step1}

\item {\em Additions / LS.} Compute $T_\dett$ and the LS estimate using it:%
\label{addls}
\bea
\label{addstep}
T_\dett \se T \cup \{i \in T^c: |(\xhat_{t,modcs})_i| > \alpha_{\dett} \} \ \ \ \  \\
(\xhat_{t,\dett})_{T_\dett} \se {A_{T_\dett}}^\dag y_t, \ \ (\xhat_{t,\dett})_{T_\dett^c} = 0
\label{lsstep}
\eea

\item {\em Deletions / LS.} Compute $\tT$ and LS estimate using it:%
\label{delete}
\bea
\label{deletestep}
\tT \se  T_{\dett} \setminus  \{i \in T_\dett: |(\xhat_{t,\dett})_i| \le \alpha_{\del} \}  \\
(\xhat_{t})_{\tT} \se {A_{\tT}}^\dag  y_t, \ \ (\xhat_{t})_{\tT^c} = 0  
\label{finalls}
\eea

\item Set $\Nhat_t = \tT$. Feedback $\Nhat_t$.  Output either $\xhat_t$ or $\xhat_{t,modcs}$.

\een
\label{modcsalgo_2}
\end{algorithm}

\bd [Define $T_{\dett,t},\Delta_{\dett,t}, \Delta_{e,\dett,t}$]
The set $T_{\dett,t}$ is the support estimate obtained after the support addition step. It is defined in (\ref{addstep}) in Algorithm \ref{modcsalgo_2}. The set $\Delta_{\dett,t}:=N_t \setminus  T_{\dett,t}$ denotes the set of missing elements from $T_{\dett,t}$ and the set $\Delta_{e,\dett,t}:=T_{\dett,t} \setminus N_t$ denotes the set of extras in it. We remove the subscript $t$ where not needed.
\label{defdett}
\ed

\subsection{Stability result for Modified-CS with Add-LS-Del}
\label{stab_modcs}

The first step to show stability is to find sufficient conditions for (a) a certain set of large coefficients to definitely get detected, and (b) to definitely not get falsely deleted, and (c) for the zero coefficients in $T_\dett$ to definitely get deleted. These can be obtained using  Corollary \ref{modcs_cs_bnd} and the following simple facts which we state as a proposition, in order to easily refer to them later. In the proposition and the three lemmas below, we remove the subscript $t$ for ease of notation.

\begin{proposition}[simple facts] Consider Algorithm \ref{modcsalgo_2}.
\ben
\item An $i \in \Delta$ will definitely get detected in step \ref{addls} if $|x_i|  > \alpha_{\dett} + \|x - \xhat_{modcs}\|$. This follows since  $ \|x - \xhat_{modcs}\| \ge \|x - \xhat_{modcs}\|_\infty \ge |(x - \xhat_{modcs})_{i}|$.
\label{det1}

\item Similarly, an $i \in T_\dett$ will definitely not get falsely deleted in step \ref{delete} if $|x_i| >  \alpha_{\del} + \|(x - \xhat_\dett)_{T_\dett}\|$.
\label{nofalsedel1}

\item All $i \in \Delta_{e,\dett}$ (the zero elements of $T_\dett$) will definitely get deleted if $\alpha_{\del} \ge \|(x - \xhat_\dett)_{T_\dett}\|$.
\label{truedel1}

\item Consider LS estimation on the known part of support $T$, i.e. consider the estimate $(\xhat_{LS})_T = {A_T}^\dag y$ and $(\xhat_{LS})_{T^c} = 0$ computed from $y:=Ax+w$. Let $\Delta = N \setminus T$ where $N$ is the support of $x$. If $\|w\| \le \eps$ and if $\delta_{|T|} < 1/2$, then $\|(x - \xhat_{LS})_{T}\| \le \sqrt{2} \eps + 2{\theta_{|T|,|\Delta|}} \|x_{\Delta}\|$. This bound is derived in \cite[equation (15)]{just_lscs}  \footnote{Instead of $\delta_{|T|} < 1/2$, one can pick any $b<1$ and the constants in the bound will change appropriately.}.
\label{errls1}

\een
\label{prop1}
\end{proposition}

Combining the above facts with Corollary \ref{modcs_cs_bnd}, we can easily get the following three lemmas.

\begin{lemma}[Detection condition]
Let $x$ be a sparse vector with support $N$ and let $y:=Ax+w$ with $\|w\| \le \eps$. Also, let $\Delta:=N \setminus T$ and $\Delta_e:=T \setminus N$.
\\ Assume that $|N| = S_N$, $|\Delta_e| \le S_{\Delta e}$, $|\Delta| \le S_\Delta$.
\\
Consider Algorithm \ref{modcsalgo_2}. For a given $b_1$, let
$$L:=\{i \in \Delta: |x_i| \ge b_1 \}.$$
All elements of $L$ will get detected in step \ref{addls} if
\ben
\item  $\delta_{S_N + S_{\Delta e} + S_\Delta} < (\sqrt{2}-1)/2$ and $S_\Delta \le S_N/3$, and
\item $b_1 > \alpha_{\dett} + 8.79\eps$.
\een
\label{detectcond_modcs}
\end{lemma}

{\em Proof: } This lemma follows from fact \ref{det1} of Proposition \ref{prop1} and Corollary \ref{modcs_cs_bnd}.

\begin{lemma}[No false deletion condition]
Let $x$ be a sparse vector with support $N$ and let $y:=Ax+w$ with  $\|w\| \le \eps$. Also, let $T_\dett, \Delta_\dett, \Delta_{e,\dett}$ be as defined in Definition \ref{defdett}.
\\ Assume that $|T_\dett| \le S_T$ and $|\Delta_\dett| \le S_\Delta$.
\\
Consider Algorithm \ref{modcsalgo_2}.  For a given $b_1$, let
$$L:=\{i \in T_\dett: |x_i| \ge b_1\}.$$
No element of $L$ will get (falsely) deleted in step \ref{delete} if
\ben
\item $\delta_{S_T} < 1/2$ and
\item $b_1 > \alpha_{\del} + \sqrt{2} \eps + 2{\theta_{S_T,S_\Delta}} \|x_{\Delta_\dett}\|$.
\een
\label{nofalsedelscond}
\end{lemma}

{\em Proof: } This lemma follows directly from fact \ref{nofalsedel1} and fact \ref{errls1} (applied with $T \equiv T_\dett$ and $\Delta \equiv \Delta_\dett$) of Proposition \ref{prop1}.

\begin{lemma}[Deletion condition]
Let $x$ be a sparse vector with support $N$ and let $y:=Ax+w$ with  $\|w\| \le \eps$. Also, let $T_\dett, \Delta_\dett, \Delta_{e,\dett}$ be as defined in Definition \ref{defdett}.
\\ Assume that $|T_\dett| \le S_T$ and $|\Delta_\dett| \le S_\Delta$.
\\
Consider Algorithm \ref{modcsalgo_2}. All elements of $\Delta_{e,\dett}$ will get deleted in step \ref{delete} if
\ben
\item $\delta_{S_T} < 1/2$ and
\item $\alpha_{\del} \ge \sqrt{2} \eps + 2{\theta_{S_T,S_\Delta}} \|x_{\Delta_\dett}\|$.
\een
\label{truedelscond}
\end{lemma}

{\em Proof: } This lemma follows directly from fact \ref{truedel1} and fact \ref{errls1} (applied with $T \equiv T_\dett$ and $\Delta \equiv \Delta_\dett$) of Proposition \ref{prop1}.

Using the above lemmas and the signal model, we can obtain sufficient conditions to ensure that, for some $d_0 \le d$, at each time $t$, $\tDelta_t \subseteq \Sset_t(d_0)$ (so that $|\tDelta_t| \le (2d_0-2)S_a$) and $|\tDelta_{e,t}|=0$, i.e. only elements smaller than $d_0 r$ may be missed and there are no extras. For notational simplicity, we state the special case below which uses $d_0=2$. The general case is given in Appendix \ref{stabres_modcs_gen} in Corollary \ref{gencase}. In fact, this is the generalized version of Corollary \ref{cor2_relax} which relaxes some assumptions of the result below.

\begin{theorem}[Stability of modified-CS with add-LS-del]
Assume Signal Model \ref{sigmod2} on $x_t$. Also assume that $y_t$ satisfies (\ref{obsmod}) with $\|w_t\| \le \eps$.
Consider Algorithm \ref{modcsalgo_2}. If 
\ben
\item {\em (addition and deletion thresholds) }
\ben
\item $\alpha_{\dett}$ is large enough so that there are at most $S_a$ false additions per unit time,
\label{addthresh}

\item $\alpha_{\del}  = \sqrt{2} \eps +  2 \sqrt{S_a} \theta_{S_0+2S_a,S_a} r $,
\label{delthresh}
\een
\label{add_del_thresh}

\item {\em (support size, support change rate)} $S_0$, $S_a$ satisfy
\ben
\item $\delta_{S_0 + 3S_a} < (\sqrt{2}-1)/2$ and $S_a \le S_0/6$, and
\label{delta_ass_2}
\item $\theta_{S_0+2S_a,S_a} < \frac{1}{2}  \frac{1}{2\sqrt{S_a}}$, 
\label{theta_ass}
\een
\label{measmodel}

\item {\em (new element increase rate) } $r \ge \max(G_1,G_2)$, where
\label{add_del}
\bea
G_1 \sdefn  \frac{ \alpha_{\dett} + 8.79 \eps }{2} \nn \\ 
G_2 \sdefn \frac{\sqrt{2} \eps}{1 - 2\sqrt{S_a }\theta_{S_0+2S_a,S_a}}  
\eea
\item {\em (initial time)} at $t=0$, $n_0$ is large enough to ensure that $\tDelta_0  \subseteq \Sset_0(2)$, $|\tDelta_0| \le 2S_a$,  $|\tDelta_{e,0}| =0$, and $|\tT_0| \le S_0$,
\label{initass}
\een
then,  
\ben

\item at all $t \ge 0$, $|\tT_t| \le S_0$, $|\tDelta_{e,t}| =0$, $\tDelta_t \subseteq \Sset_t(2)$ and so $|\tDelta_t| \le 2S_a$,
\item at all $t > 0$, $|T_t| \le S_0$, $|\Delta_{e,t}| \le S_a$, and $|\Delta_t| \le 2S_a$,

\item at all $t > 0$,  $|\tT_{\dett,t}| \le S_0+2S_a$, $|\tDelta_{e,\dett,t}| \le 2S_a$, and $|\tDelta_{\dett,t}| \le S_a$ 

\item at all $t > 0$,  $\|x_t-\xhat_t\| \le \sqrt{2} \eps + (2\theta_{S_0,2S_a}+1) \sqrt{2S_a} r$

\item at all $t > 0$,  $\|x_t - \xhat_{t,modcs}\| \le 8.79 \eps$ 
\een
\label{stabres_modcs}
\end{theorem}

{\em Proof: } The complete proof is given in Appendix \ref{proof_addLSdel_modcs}. This proof also follows by induction, but is more complicated than that of Theorem \ref{stabres_simple_modcs}. 
The induction step consists of three parts.
\bi
\item First, we use the induction assumption; the fact that $T_t = \tT_{t-1} = \Nhat_{t-1}$; and the fact that $N_t = N_{t-1} \cup \Aset_t \setminus \Rset_t$ to bound $|T_t|,|\Delta_{e,t}|,|\Delta_t|$. This part of the proof is the same as that of Theorem \ref{stabres_simple_modcs}. The next two parts are quite different.%
\item We use the bounds from the first part; equation (\ref{sseteq_2}); Lemma \ref{detectcond_modcs}; the limit on the number of false detections from condition \ref{addthresh}; and $|T_\dett| \le |N| + |\Delta_{e,\dett}|$ to bound  $|\Delta_{\dett,t}|,|\Delta_{e,\dett,t}|,|T_{\dett,t}|$.
\item Finally, we use the bounds from the second part; Lemmas \ref{nofalsedelscond} and \ref{truedelscond}; and $|\tT| \le |N|+|\tDelta_e|$ to bound $|\tDelta_t|,|\tDelta_{e,t}|,|\tT_t|$.
\ei

\subsection{Discussion} 
\label{modcsald_discuss}
Notice that condition \ref{theta_ass} may become difficult to satisfy as soon as $S_a$ increases, which will happen when the problem dimension, $m$, increases, and consequently $S_0$ increases, even though $S_a$ and $S_0$ remain small fractions of $m$, e.g. typically $S_0 \approx 10\%m$ and $S_a \approx 2\% \text{ - } 10\% S_0 \approx 0.2\% \text{ - } 1\% m$. The reason we get this condition is because in facts \ref{nofalsedel1} and \ref{truedel1} of Proposition \ref{prop1}, and hence also in Lemmas \ref{nofalsedelscond} and \ref{truedelscond} and in the final result, we bound the $\ell_\infty$ norm of the LS step error, $(x - \xhat_\dett)_{T_\dett}$, by its $\ell_2$ norm. This is clearly a loose bound. It holds with equality only when the entire LS step error is concentrated in one dimension.

In practice, as observed in our simulations, the LS step error is actually quite spread out, since the LS step tends to reduce the bias in the estimate, at least as long as the number of misses in $T_\dett$ is small and $A_{T_\dett}$ is well conditioned (which are required conditions for stability anyway and are enforced by conditions \ref{add_del} and \ref{delta_ass_2} of Theorem \ref{stabres_modcs}).
Thus, it is not unreasonable to assume that $\|(x - \xhat_\dett)_{T_\dett}\|_\infty \le C \|(x - \xhat_\dett)_{T_\dett}\|$ for some $C < 1$. From simulations, it is observed that $C = \frac{\zeta_m}{\sqrt{S_a}}$ works. Here $\zeta_m$ is slightly more than one and increases very slowly with $m$, e.g. for $m=200$, $\zeta_m = 1.11$, for $m=1000$, $\zeta_m = 1.23$ and for $m=2000$, $\zeta_m = 1.38$. The above numbers were obtained when we simulated according to the generative model for Signal Model \ref{sigmod2} given in Appendix \ref{generativemodel1}; we used $S_0=0.1m$, $S_a = 0.01m$, and $r=1$; the matrix $A$ was random Gaussian, with $n = 0.3861 S_0 \log_2 m$; the noise, $w_t$, was independent identically distributed (i.i.d.) $uniform(-c,c)$ in various dimensions and over time and we used $c=0.1266$; and we set $\alpha_\dett=c/2$ and $\alpha_\del=r/2$ \footnote{We computed $\zeta_m$ by computing the maximum of $\frac{\|(x_t - \xhat_{\dett,t})_{T_{\dett,t}}\|_\infty \sqrt{S_a}}{  \|(x_t - \xhat_{\dett,t})_{T_{\dett,t}}\|}$ over time and over 500 independent simulations for $m=200$ (and over 50 for $m=1000,2000$). The matrix $A$ was chosen once and fixed. We sampled over the distributions of $w_t$ and $x_t$.}. Similar conclusions were obtained for $r=3/4$ and $2/3$.

With using the extra assumption $\|(x - \xhat_\dett)_{T_\dett}\|_\infty \le \frac{\zeta_m}{\sqrt{S_a}} \|(x - \xhat_\dett)_{T_\dett}\|$ in facts \ref{nofalsedel1} and \ref{truedel1} of Proposition \ref{prop1}, Lemmas \ref{nofalsedelscond} and \ref{truedelscond} get replaced by the following two lemmas. With using these new lemmas, condition \ref{theta_ass} of Theorem \ref{stabres_modcs} will get replaced by $\theta_{S_0+2S_a,S_a} < \frac{1}{4 \zeta_m}$ which is an easily satisfiable condition. Moreover, this also makes the lower bound on the required value of $r$ (rate of coefficient increase/decrease) smaller.

\begin{lemma}[No false deletion condition -- weaker]
Let $x$ be a sparse vector with support $N$ and let $y:=Ax+w$, with $\|w\| \le \eps$. Also, let $T_\dett, \Delta_\dett, \Delta_{e,\dett}$ be as defined in Definition \ref{defdett}.
\\ Assume that $|T_\dett| \le S_T$ and $|\Delta_\dett| \le S_\Delta$.
\\ Consider Algorithm \ref{modcsalgo_2}. Assume that the LS step error is spread out enough to ensure that
$$\|(x - \xhat_\dett)_{T_\dett}\|_\infty \le \frac{\zeta_m}{\sqrt{S_a}} \|(x - \xhat_\dett)_{T_\dett}\|.$$
For a given $b_1$, let
$$L:=\{i \in T_\dett: |x_i| \ge b_1\}.$$
No elements of $L$ will get (falsely) deleted in step \ref{delete} if
\ben
\item $\delta_{S_T} < 1/2$, and
\item $b_1 > \alpha_{\del} + \frac{\zeta_m}{\sqrt{S_a}} (\sqrt{2} \eps + 2{\theta_{S_T,S_\Delta}} \|x_{\Delta_\dett}\|)$.%
\een
\label{nofalsedels_equalLS}
\end{lemma}
%

\begin{lemma}[Deletion condition -- weaker]
Let $x$ be a sparse vector with support $N$ and let $y:=Ax+w$, with $\|w\| \le \eps$. Also, let $T_\dett, \Delta_\dett, \Delta_{e,\dett}$ be as defined in Definition \ref{defdett}.
\\ Assume that $|T_\dett| \le S_T$ and $|\Delta_\dett| \le S_\Delta$.
\\
Consider Algorithm \ref{modcsalgo_2}. Assume that the LS step error is spread out enough to ensure that
$$\|(x - \xhat_\dett)_{T_\dett}\|_\infty \le \frac{\zeta_m}{\sqrt{S_a}} \|(x - \xhat_\dett)_{T_\dett}\|.$$
All elements of $\Delta_{e,\dett}$ will get deleted in step \ref{delete} if
\ben
\item $\delta_{S_T} < 1/2$ and
\item $\alpha_{\del} \ge  \frac{\zeta_m}{\sqrt{S_a}}(\sqrt{2} \eps + 2{\theta_{S_T,S_\Delta}} \|x_{\Delta_\dett}\|)$.
\een
\label{truedelscond_equalLS}
\end{lemma}
%

By using Lemmas \ref{nofalsedels_equalLS} and \ref{truedelscond_equalLS} instead of Lemmas \ref{nofalsedelscond} and \ref{truedelscond} respectively,  and doing everything else exactly as in the proof of Theorem \ref{stabres_modcs}, we get the following corollary. 


\begin{corollary}[Stability of modified-CS with add-LS-del -- 2]
Assume Signal Model \ref{sigmod2} on $x_t$. Also assume that $y_t$ satisfies (\ref{obsmod}) with $\|w_t\| \le \eps$.
Let $$e_t:=(x_t - \xhat_{\dett,t})_{T_{\dett,t}}$$ denote the LS step error. Assume that the LS step error is spread out enough so that
\bea
\|e_t\|_\infty \le \frac{\zeta_m}{\sqrt{S_a}} \|e_t\|
\label{linf_l2}
\eea
at all times, $t$. Consider Algorithm \ref{modcsalgo_2}. If 
\ben
\item {\em (addition and deletion thresholds) }
\ben
\item $\alpha_{\dett}$ is large enough so that there are at most $S_a$ false additions per unit time,
\label{addthresh}

\item $\alpha_{\del}  = \sqrt{\frac{2}{S_a}}\zeta_m \eps + 2 \theta_{S_0+2S_a,S_a} \zeta_m r $, 
\label{delthresh}
\een
\label{add_del_thresh}

\item {\em (support size, support change rate)} $S_0$, $S_a$ satisfy
\ben
\item  $\delta_{S_0 + 3S_a} < (\sqrt{2}-1)/2$ and $S_a \le S_0/6$, and
\label{measmod_delta}

\item $\theta_{S_0+2S_a,S_a} <  \frac{1}{4 \zeta_m}$
\label{measmod_theta}
\een
\label{measmodel} 

\item {\em (new element increase rate) } $r \ge \max(G_1,G_2)$, where
\label{add_del}
\bea
G_1 \sdefn  \frac{ \alpha_{\dett} + 8.79 \eps }{2}  \nn \\ 
G_2 \sdefn \frac{\sqrt{2} \zeta_m \eps}{\sqrt{S_a} (1 - 2\theta_{S_0+2S_a,S_a} \zeta_m )} 
\eea

\item {\em (initial time)} at $t=0$, $n_0$ is large enough to ensure that $\tDelta  \subseteq \Sset_0(2)$, $|\tDelta| \le 2S_a$,  $|\tDelta_e| =0$, $|\tT| \le S_0$,
\label{initass}

\een
then all conclusions of Theorem  \ref{stabres_modcs} hold.
\label{cor2_relax}
\end{corollary}



A generalization of the above corollary, that allows the support error to stabilize at $(2d_0-2)S_a$, for some $d_0 \le d$, is given in Appendix \ref{stabres_modcs_gen} in Corollary \ref{gencase}.

Recall that $\zeta_m/\sqrt{S_a}$ is smaller than one. For example, in our simulations, when $m=2000$, $\zeta_m=1.38$, while $\sqrt{S_a}=\sqrt{20}=4.47$. Also, $\zeta_m$ increases very slowly with $m$ (slower than $O(\log m)$) where as $\sqrt{S_a}$ typically increases as $\sqrt{m}$. Thus, conditions \ref{delthresh}, \ref{measmod_theta} and \ref{add_del} are significantly weaker compared to those in Theorem \ref{stabres_modcs}, while others are the same. In particular, now condition \ref{measmod_theta} is easy to satisfy.


Let us compare this result with that for modified-CS given in Theorem \ref{stabres_simple_modcs}. Consider the lower bound on $r$ required by both results. In the above result, since $\theta_{S_0+2S_a,S_a} < 1/(4\zeta_m)$, so $G_2 < \frac{2\sqrt{2} \zeta_m}{\sqrt{S_a}} \eps < 2.9 \eps < \frac{8.79 \eps}{2} < G_1$ and thus $G_1$ is what decides the minimum allowed value of $r$.
Because of add-LS-del, the addition threshold, $\alpha_{\dett}$, can now be much smaller, as long as the number of false adds is small\footnote{e.g. in simulations with $m=200$, $S_0=20$, $S_a=2$,  $r=1$ (or even for $r=2/3$), $n=59$, $(w_t)_j \sim^{i.i.d.} uniform(-c,c)$ with $c=0.1266$, and $\alpha_{\del}=r/2$, we were able to use $\alpha_{\dett}=c/2=0.06$ and still ensure that the number of false adds is less than or equal to $S_a$ (details in Sec. \ref{sims}).}. If $\alpha_{\dett}$ is close to zero, the value of $G_1$ is almost half that of $G$ in Theorem \ref{stabres_simple_modcs}. Thus the minimum coefficient increase rate, $r$, required by the above result is almost half of that required by Theorem \ref{stabres_simple_modcs}.
On the other hand, the above result also requires condition \ref{measmod_theta} on $\theta$ which Theorem \ref{stabres_simple_modcs} does not, but this condition is typically weaker than condition \ref{measmod_delta} since $\theta_{S_0+2S_a,S_a}$ is smaller than $\delta_{S_0+3S_a}$ where as the right hand sides do not differ by much.

The above is also demonstrated in Figs. \ref{simfig_n59_r1} and \ref{simfig_n59_r2b3}. For $r=1$, both are stable, but for $r=2/3$, modified-CS is unstable while modified-CS with add-LS-del is still stable.

Finally, let us compare our result with the simple CS result given in Corollary \ref{cs_bnd}. Corollary \ref{cs_bnd} needs $\delta_{2S_0} < (\sqrt{2}-1)/2=0.207$ to achieve the same error bound as our result. On the other hand, if the LS step error is spread out enough, we only need $\delta_{S_0 + 3S_a} < (\sqrt{2}-1)/2=0.207$ and $\theta_{S_0+2S_a,S_a} < 1/(4 \zeta_m)$.
When $S_a \ll S_0$ (slow support change), the first condition is clearly weaker than what CS needs. The second condition is also weaker since $\theta_{S_0+2S_a,S_a}$ is significantly smaller than $\delta_{2S_0}$ where as the right hand sides $0.207$ and $0.25/\zeta_m$ are roughly equal. A quantitative comparison can be done by using the upper bounds $\theta_{u,k} \le \delta_{u+k}$ \cite{decodinglp} and $\delta_{ck} \le c \delta_{2k}$ \cite{cosamp}.
If $S_a = 0.02 S_0$, then $\delta_{2S_0} = \delta_{100S_a} \le 100 \delta_{2S_a}$ and $\theta_{S_0+2S_a,S_a} \le \delta_{S_0+3S_a} \le 53\delta_{2S_a}$. Thus, the CS condition is stronger as long as $\zeta_m < (100/53) (0.25/0.207) = 2.28$. If $S_a = 0.1 S_0$, then the CS condition is stronger if $\zeta_m < 1.9$.





\begin{remark}
In the discussion so far we have used the special case stability results where we find conditions to ensure that the misses remain below $2S_a$.
Let us look at the general form of the result -- Corollary \ref{gencase} in Appendix \ref{stabres_modcs_gen} -- where we provide conditions to ensure that,  for some $d_0 \le d$, the misses are below $(2d_0-2)S_a$. In Corollary \ref{gencase}, using an argument similar to the one above, $\check{G}_2 < \check{G}_1$ holds for any $d_0$. Also, notice that, if the rate of coefficient increase, $r$, is smaller, $r \ge \check{G}_1$ will hold for a larger value of $d_0$. This means that the support error bound,  $(2d_0-2)S_a$, will be larger. This, in turn, decides what conditions on $\delta$ and $\theta$ are needed (in other words, how many measurements, $n$, are needed). Smaller $r$ means a larger $d_0$ is needed which, in turn, means that stronger conditions on $\delta, \theta$ (larger $n$) are needed.
Thus, for a given $n$, as $r$ is reduced, the algorithm will stabilize to larger and larger support error levels (larger $d_0$) and finally become unstable (because the given $n$ does not satisfy the conditions on $\delta,\theta$ for the larger $d_0$).
%
\end{remark}

The above is demonstrated empirically in Fig. \ref{fig1}. The last three rows of this figure used $n=59$. When $r=1$,  modified-CS with add-LS-del is stable at zero support errors. When $r$ is reduced to $2/3$, it is stable at mean support errors less than 0.3\%. When $r$ is reduced to $2/5$ it becomes unstable.%




\section{Stability of LS-CS}
\label{addLSdel_lscs}
In \cite{just_lscs,kfcspap}, we introduced Least Squares CS-residual (LS-CS) as one of the first solutions to the problem of recursively reconstructing sparse signal sequences with slow time-varying sparsity patterns. We summarize the complete LS-CS algorithm in Algorithm \ref{lscsalgo_2}. LS-CS uses partial knowledge of support, $T$, in a different way than modified-CS. It first computes an initial LS estimate on the set $T$, as in (\ref{initlsstep}), and then computes the observation residual, as in (\ref{deftty0}). Noisy CS is done on this observation residual, as in (\ref{simplecs}), and the solution is added back to the initial LS estimate, as in (\ref{xhatcsres}). The add-LS-del approach described earlier is used for support estimation.


\begin{algorithm}[h!]
\caption{{\bf \small Least Squares CS-residual (LS-CS)}} 
For $t\ge 0$, do
\ben
\item {\em Simple CS. } Do as in Algorithm \ref{modcsalgo_2}.

\item {\em CS-residual. }
\label{csresstep}
\ben
\item Use $T:=\Nhat_{t-1}$ to compute the initial LS estimate, $\xhat_{t,\text{init}}$, and the LS residual, $\tty$, as follows.
\label{initls}
\bea
\label{initlsstep}
(\xhat_{t,\text{init}})_{T} \se {A_{T}}^\dag y_t, \ \ (\xhat_{t,\text{init}})_{T^c} = 0  \\
\tty \se  y_t - A \xhat_{t,\text{init}}  
\label{deftty0}
\eea

\item Do noisy CS on the LS residual, i.e. solve
\bea
\min_\beta \|\beta\|_1 \ s.t. \ \| \tty - A \beta \| \le \eps
\label{simplecs}
\eea
and denote its output by $\betahat_t$. Compute 
\bea
\xhat_{t,\CSres} : = \betahat_t + \xhat_{t,\text{init}}.
\label{xhatcsres}
\eea
\een

\item {\em Additions / LS.} Compute $T_\dett$ and the LS estimate on it as in  Algorithm \ref{modcsalgo_2}. Use $\xhat_{t,\CSres}$ instead of $\xhat_{t,modcs}$ for estimating $T_\dett$.
\label{addstep_lscs}

\item {\em Deletions / LS.} Compute $\tT$ and the LS estimate on it as in Algorithm \ref{modcsalgo_2}.

\item Set $\Nhat_t = \tT$. Output $\xhat_t$. Feedback $\Nhat_t$.

\een
\label{lscsalgo_2}
\end{algorithm}

The CS-residual step error, $x_t - \xhat_{t,\CSres}$, where $\xhat_{t,\CSres}$ is defined in (\ref{xhatcsres}),  can be bounded as follows. The proof is easy and follows in the same way as that for \cite[Corollary 1]{just_lscs} where noisy CS is done using Dantzig selector instead of (\ref{simplecs}). 
We use (\ref{simplecs}) here to keep the comparison with modified-CS easier.

\begin{lemma}[CS-residual error bound \cite{just_lscs}]
Let $x$ be a sparse vector with support $N$ and let $y:=Ax+w$ with $\|w\| \le \eps$. Also, let $\Delta:=N \setminus T$ and $\Delta_e:=T \setminus N$. Consider step \ref{csresstep} of Algorithm \ref{lscsalgo_2}.
If
\bi
\item $\delta_{2|\Delta|} < (\sqrt{2}-1)/2$ and
\item $\delta_{|T|} < 1/2$,
\ei
then
\bea
&& \|x - \xhat_{\CSres}\| \le C' \eps +  \theta_{|T|,|\Delta|} C'' \|x_\Delta\|, \ \text{where} \nn \\
&& C'  \equiv  C'(|T|,|\Delta|)  \defn C_1(2|\Delta|) + \sqrt{2} C_2(2|\Delta|) \sqrt{\frac{|T|}{|\Delta|}}, \nn \\
&& C''  \equiv  C''(|T|,|\Delta|) \defn 2 C_2(2|\Delta|) \sqrt{\frac{|T|}{|\Delta|}}, \nn \\  
&& \text{$C_1(S)$ is defined in (\ref{defC1s}),~}  C_2(S) \defn 2\frac{1 + (\sqrt{2}-1) \delta_S}{1 - (\sqrt{2}+1) \delta_S} \ \ \ \ \
\eea
\label{lscs_bnd}
\end{lemma}


\subsection{Stability result for LS-CS}
Our overall approach is similar to the one discussed in the previous section for modified-CS with add-LS-del. The key difference is in the detection condition lemma, which we give below. Its proof is given in Appendix \ref{proof_detectcond_lscs}. This lemma is different from Lemma \ref{detectcond_modcs} because, unlike modified-CS, the CS-residual error bound at time $t$ also depends on the magnitudes of the elements in the initial missed set $\Delta_t$.

\begin{lemma}[Detection condition for LS-CS]
Let $x$ be a sparse vector with support $N$ and let $y:=Ax+w$ with $\|w\| \le \eps$. Also, let $\Delta:=N \setminus T$ and $\Delta_e:=T \setminus N$. Assume that $|T| \le S_T$ and $|\Delta| \le S_\Delta$. Assume that $\|x_\Delta\|_\infty \le b$.
\\
Consider step \ref{addstep_lscs} of Algorithm \ref{lscsalgo_2}.
For a $\gamma \le 1$, let $$L_1:=\{i \in \Delta: \gamma b \le |x_i| \le b \}$$ and let $$L_2:= \Delta \setminus L_1 = \{i \in \Delta:  |x_i| < \gamma b \}.$$
Assume that $|L_1| \le S_{L 1}$ and $\|x_{L_2}\| \le \kappa b$.
All $i \in L_1$ 
will definitely get detected at the current time if
\ben
\item $\delta_{2S_\Delta} < (\sqrt{2}-1)/2$,
\item $\delta_{S_T} < 1/2$,
\item $\max_{|\Delta| \le S_\Delta}  \theta_{S_T,|\Delta|} C''(S_T,|\Delta|) \le \frac{\gamma}{2(\sqrt{S_{L 1}}+\kappa)}$, and
\item
\bea
&& \max_{|\Delta| \le S_\Delta} \frac{ \alpha_{\dett} + C'(S_T,|\Delta|)\eps }{\gamma - {\theta_{S_T,|\Delta|}} C''(S_T,|\Delta|) (\sqrt{S_{L 1}} + \kappa)} < b \ \  \nn
\label{detcond_lscs}
\eea
where $C'(.,.)$, $C''(.,.)$ are defined in Lemma \ref{lscs_bnd}.
\een
\label{detectcond_lscs}
\end{lemma}

{\em Proof: } The proof is given in Appendix \ref{proof_detectcond_lscs}.

The stability result then follows in the same fashion as Theorem \ref{stabres_modcs}. The only difference is that instead of Lemma \ref{detectcond_modcs}, we apply Lemma \ref{detectcond_lscs} with $S_T=S_0$, $S_\Delta=2S_a$, $b=2r$, $\gamma=1$, $S_{L 1}=S_a$ and $\kappa = \frac{\sqrt{S_a} r}{2r} = \frac{\sqrt{S_a}}{2}$.

\begin{theorem}[Stability of LS-CS]
Assume Signal Model \ref{sigmod2} on $x_t$. Also assume that $y_t$ satisfies (\ref{obsmod}) with $\|w_t\| \le \eps$.
Consider Algorithm \ref{lscsalgo_2}. If 
\ben
\item  {\em (addition and deletion thresholds) }
\ben
\item $\alpha_\dett$ is large enough so that there are at most $S_a$ false additions per unit time,
\label{addthresh_lscs}

\item $\alpha_\del  = \sqrt{2} \eps + 2 \sqrt{S_a} \theta_{S_0+2S_a,S_a} r $
\label{delthresh_lscs}
\een

\item {\em (support size, support change rate)} $S_0, S_a$ satisfy
\label{measmodel_lscs}
\ben
\item $\delta_{4S_a} < (\sqrt{2}-1)/2$
\label{measmodel_lscs_1a}
\item $\delta_{S_0+2S_a} < 1/2$
\label{measmodel_lscs_1b}

\item $\max_{|\Delta| \le 2S_a} \theta_{S_0,|\Delta|} C''(S_0,|\Delta|)   <  \frac{1}{3 \sqrt{S_a}}$
\label{measmodel_lscs_2a}

\item $\theta_{S_0+2S_a,S_a} < \frac{1}{2} \frac{1}{2 \sqrt{S_a} }$
\label{measmodel_lscs_2b}

\een

\item {\em  (new element increase rate) } $ r \ge \max(\tilde{G}_1,\tilde{G}_2)$, where
\label{add_del_lscs}
\bea
&& \tilde{G}_1 \defn \max_{|\Delta| \le 2S_a}  [\frac{ \alpha_\dett + C'(S_0,|\Delta|)\eps  }{2 -  3 {\theta_{S_0,|\Delta|}} \sqrt{S_a} C''(S_0,|\Delta|)}] \nn \\
%
&& \tilde{G}_2 \defn \frac{\sqrt{2} \eps}{1 - 2\sqrt{S_a}\theta_{S_0+2S_a,S_a}}
\eea

\item {\em (initialization) }  (same condition as in Theorem \ref{stabres_modcs})
\een
then, all conclusions of Theorem \ref{stabres_modcs} hold for LS-CS, except the last one. This is replaced by $\|x_t - \xhat_{t,\CSres}\| \le \max_{|\Delta| \le 2S_a} [ C'(S_0,|\Delta|)\eps  + (\theta_{S_0,|\Delta|} C''(S_0,|\Delta|) + 1) \sqrt{2S_a} r]$.
\label{stabres_lscs}
\end{theorem}

\subsection{Discussion}
Notice that conditions \ref{measmodel_lscs_2a} and \ref{measmodel_lscs_2b} are the difficult conditions to satisfy as the problem size, $m$, increases and consequently $S_0$ and $S_a$ increase. We get condition \ref{measmodel_lscs_2b} because we bound the $\ell_\infty$ norm of the addition LS step error by its $\ell_2$ norm. This can be relaxed to $\theta_{S_0+2S_a,S_a} < 1/(4 \zeta_m)$ in the same fashion as in the previous section.

Consider condition \ref{measmodel_lscs_2a}. We get this condition because (i) we upper bound the $\ell_\infty$ norm of the CS-residual step error, $x_t - \xhat_{t,\CSres}$, by its $\ell_2$ norm in Lemma \ref{detectcond_lscs}; and (ii) in the proof of Lemma \ref{lscs_bnd}, we upper bound the $\ell_1$ norm of the initial LS step error, $(x_t - \xhat_{t,\text{init}})_T$, by $\sqrt{|T|}$ times its $\ell_2$ norm (this results in the expression for $C''$ given in Lemma \ref{lscs_bnd}). If we can relax (i), we can try to weaken the required condition, but it will still be stronger than what modified-CS with add-LS-del or modified-CS need.  For example, if we can assume a bound similar to (\ref{linf_l2}) for the CS-residual step error, and if additionally, we assume that, in the range $|\Delta| \le 2S_a$, $\theta_{S_0,|\Delta|}C''(S_0,|\Delta|)$ is largest for $|\Delta|=2S_a$, condition \ref{measmodel_lscs_2a} will get relaxed to something like $\theta_{S_0,2S_a} C_2(2S_a) \le \frac{1}{3 \zeta_m}\sqrt{\frac{S_a}{2S_0}}$. This is still stronger than condition \ref{measmod_theta} of Corollary \ref{cor2_relax}, primarily because of $\sqrt{S_a/S_0}$.

The above is also observed in our simulations. In Fig. \ref{fig1}, LS-CS needs a larger $n$ ($n=65$) for stability where as for modified-CS with add-LS-del or modified-CS, $n=59$ suffices. We show the results for $r=1$ or lower, but even when we increased $r$ to $r=2$ or $r=3$, LS-CS was still unstable with $n=59$. The simulation details are given in Sec.~\ref{sims}.%

\section{Simulation Results}
\label{sims}
We compared modified-CS (mod-CS), as given in Algorithm \ref{modcsalgo}, modified-CS with Add-LS-Del (mod-CS-add-LS-del), as given in Algorithm \ref{modcsalgo_2} (with final output $\xhat_t$), LS-CS, as given in Algorithm \ref{lscsalgo_2}, and simple CS for a few different choices of $n$ and $r$. The results are shown in Fig. \ref{fig1} where we show four rows of plots. In each row, we plot the normalized mean squared error (NMSE), $\frac{\E[\|x_t - \xhat_t\|^2]}{\E[\|x_t\|^2]}$, the normalized mean extras, $\frac{\E[|\Nhat_t \setminus N_t|]}{\E[|N_t|]}$, and the normalized mean misses, $\frac{\E[| N_t \setminus \Nhat_t|]}{\E[|N_t|]}$ in the left, middle and right columns respectively. Here $\E[.]$ denotes the empirical mean over the 500 realizations.

In all rows, we used the generative model for Signal Model \ref{sigmod2} from Appendix \ref{generativemodel1} with $m=200$, $S_0=20$, $S_a=2$. The measurement noise, $(w_t)_j \sim^{i.i.d.} uniform(-c,c)$ with $c=0.1266$, i.e. it was i.i.d. uniform in all dimensions and over time.
Each element of the measurement matrix, $A$, was i.i.d. zero mean random Gaussian. Fig. \ref{simfig_n65_r1} used $n=65$, $r=1$ and $d=3$, while the other three rows used $n=59$. Fig. \ref{simfig_n59_r1} used $n=59$, $r=1$, $d=3$; Fig. \ref{simfig_n59_r2b3} used $n=59$, $r=2/3$, $d=3$; and Fig. \ref{simfig_n59_r2b5} used $n=59$, $r=2/5$, $d=5$. Our simulations selected $A$ once and kept it fixed, but Monte Carlo averaged over $w_t$ and $x_t$.

We set the addition threshold, $\alpha_{\dett}$, to be at the noise level - we set it to $c/2$. Assuming that the LS step after support addition gives a fairly accurate estimate of the nonzero values, one can set the deletion threshold, $\alpha_{\del}$, to a larger value of $\alpha_{\del}=r/2$ and still ensure that there are no (or very few) false deletions. Larger deletion threshold ensures that all (or most) of the false additions and removals get deleted. Modified-CS used a single threshold, $\alpha$, somewhere in between $\alpha_{\dett}$ and $\alpha_{\del}$. We set $\alpha = ((c/2)+(r/2))/2$ (we picked this after trying a few different options for $\alpha$). Also, we did not do anything at $t=0$. We just started our simulation at $t=1$ with the assumption that $|\tDelta_0|=2$, $|\tDelta_{e,0}|=0$ and hence $|\tT_0|=S_0-2$ (i.e. the initial time condition of all our theorems holds). 

Notice, from the plots, that LS-CS needs at least $n=65$ for stability (compare Fig. \ref{simfig_n65_r1} with Fig. \ref{simfig_n59_r1}) where as mod-CS-add-LS-del and mod-CS are stable even with $n=59$. We also tried using $n=59$ and larger values of $r$, but even with $r=3$ LS-CS was still unstable in a few cases. 

Secondly, even with $n=65$, simple CS NMSE is about 20\% where as mod-CS-add-LS-del and LS-CS are stable at 0.1\% and mod-CS is stable at 0.3\%. We do not show support recovery errors for simple CS since they were very large.  With $n=59$, simple CS NMSE goes up to 30\%.
We also show the NMSE plot for simple Gauss-CS (CS followed by a final LS step on the estimated support, done in a fashion similar to Gauss-Dantzig selector \cite{dantzig}). Since the CS error itself is so large, this debiasing step does not help.

When $n=65$ and $r=1$, mod-CS is stable, but has larger error than both LS-CS and mod-CS-add-LS-del.
When $n=59$ and $r=1$, LS-CS becomes unstable. But, mod-CS and mod-CS-add-LS-del are still stable, with mod-CS being stable at a larger error (both larger support error and MSE) than mod-CS-add-LS-del. When $r$ is reduced to $2/3$, mod-CS also becomes unstable. But mod-CS-add-LS-del is still stable, though at higher error values than when $r=1$. When $r$ is further reduced to $2/5$, even mod-CS-add-LS-del becomes unstable.%

Mod-cs-add-LS-del uses a better support estimation method and thus its extras and misses are both much smaller than those of mod-CS. As a result, (a) it can remain stable for smaller values of $r$ than mod-CS; and (b) when both are stable, its reconstruction error is smaller than that of mod-CS.

\section{Conclusions}
\label{conclusions}
Under mild assumptions, we showed the ``stability" of modified-CS and its improved version, modified-CS with add-LS-del, and of LS-CS for recursive sparse signal sequence reconstruction. By ``stability" we mean that the number of misses from the current support estimate and the number of extras in it remain bounded by a time-invariant value at all times. 
Under slow support change, the results are meaningful, i.e. the bound is small compared to the support size.
A direct corollary is that the reconstruction errors are also bounded by time-invariant and small values.

We can argue that our results ensure stability under weaker assumptions that those required by simple CS. We are also able to compare the implications of the results for the three recursive algorithms and argue that modified-CS with add-LS-del needs the weakest conditions on both the number of measurements, $n$, and on the rate of coefficient magnitude increase/decrease, $r$. Modified-CS needs similar conditions on $n$, but needs $r$ to be larger. LS-CS needs the strongest conditions on both $n$ and $r$. All of our conclusions are supported by empirical performance evaluations that compare the reconstruction error as well as the support recovery errors using Monte Carlo simulations.

Two open questions that remain are as follows. The first is how to show stability for a stochastic model of signal change that models small random variations around the mean number of support additions/removals and around the mean magnitude increase/decrease rate.
A second open question is to show stability under reasonable assumptions for approaches that also use slow signal value change, e.g. KF-CS \cite{kfcsicip,just_lscs} or regularized modified-CS \cite{isitmodcs} or of \cite{schniter_track}.

\appendix
\subsection{Generative Models for  Signal Model \ref{sigmod2}}
\label{generativemodel}
To help understand Signal Model \ref{sigmod2} better, we provide here two possible generative models that satisfy its assumptions. In both cases, at $t=0$, the support size is $S_0$ and it contains $2S_a$ elements each with magnitude $r,2r, \dots (d-1)r$, and $(S_0-(2d-2)S_a)$ elements with magnitude $M$.
\subsubsection{Generative Model 1}
\label{generativemodel1}
This assumes that when a new element gets added to the support, its magnitude keeps increasing at rate $r$ until it reaches $M:=dr$. An analogous model is assumed for decrease until removal from support. The sign is selected as $+1$ or $-1$ with equal probability when the element gets added to the support, but remains the same after that.

Mathematically this can be described as follows. Let $(x_t)_i = (m_t)_i (s_t)_i$ where $(m_t)_i$ denotes the magnitude and $(s_t)_i$ denotes the sign of $(x_t)_i$ at time $t$.
\\ At any $t>0$, do the following.
\ben
\item Update 
\bea
\Iset_t(j) \se \Iset_{t-1}(j-1), \ \text{for all} \ 2 \le j \le d, \text{~and} \nn \\
\Dset_t(j) \se \Dset_{t-1}(j+1), \ \text{for all} \ 0 \le j \le d-2
\eea
where $\Iset_t(j)$ and $\Dset_t(j)$ are defined in Definition \ref{defIset}. Recall that the removed set, $\Rset_t = \Dset_t(0)$.%

\item Generate 
\ben
\item the new addition set, $\Aset_t= \Iset_t(1)$, of size $S_a$ uniformly at random from ${N_{t-1}}^c$, and
\item the new decreasing set, $\Dset_t(d-1)$, of size $S_a$ uniformly at random from $\{i \in N_{t-1}: (x_{t-1})_i = M \}$.%
\een

\item Update the coefficients' magnitudes as follows.
\bea
(m_t)_i \se \left\{ \begin{array}{ll}
              (m_{t-1})_i + r, & \ i \in \cup_{j=1}^d \Iset_t(j)  \\
              (m_{t-1})_i - r, & \ i \in \cup_{j=0}^{d-1} \Dset_t(j) \\
              (m_{t-1})_i,     & \ i \in {\cal C}_t
              \end{array}
              \right.
\eea
where ${\cal C}_t:= N_t \setminus \{\cup_{j=1}^d \Iset_t(j) \ \cup \ \cup_{j=0}^{d-1} \Dset_t(j) \}$.
\item Update the signs as follows.
\bea
(s_t)_i \se \left\{ \begin{array}{ll}
              (s_{t-1})_i, & \ i \in N_t \setminus \Aset_t  \\
              iid(\pm 1), & \ i \in \Aset_t \\
              0,     & \ i \in N_t^c
              \end{array}
              \right.
\eea
where $iid(\pm 1)$ refers to generating the sign as +1 or -1 with equal probability and doing this independently for each element $i$.

\item Set $(x_t)_i = (m_t)_i (s_t)_i$ for all $i$.
\een
Our simulations used the above model.

\subsubsection{Generative Model 2}
A second reasonable generative model selects any $S_a$ out of the $2S_a$ elements with current magnitude $jr$ and increase them, and decreases the other $S_a$ elements. In other words, it replaces the first step above by the following, while keeping the rest of the steps the same.%
\ben
\item Generate
\ben
\item $\Iset_t(j)$ of size $S_a$ uniformly at random from $\{i \in N_{t-1}: (x_{t-1})_i = (j-1)r \}$ for all $2 \le j \le d$.
\item $\Dset_t(j)$ of size $S_a$ uniformly at random from $\{i \in N_{t-1}: (x_{t-1})_i = (j+1)r \}$ for all $0 \le j \le d-2$.
\een
\een

\subsection{Appendix: Proof of Theorem \ref{stabres_simple_modcs}}
\label{proof_simple_modcs}

We prove the first claim by induction. Using condition \ref{initass_simple} of the theorem, the claim holds for $t=0$. This proves the base case. For the induction step, assume that the claim holds at $t-1$, i.e. $|\tDelta_{e,t-1}| =0$, $|\tT_{t-1}| \le S_0$, and $\tDelta_{t-1} \subseteq \Sset_{t-1}(2)$ so that $|\tDelta_{t-1}| \le 2S_a$. Using this we prove that the claim holds at $t$. In the proof, we use the following facts often: (a) $\Rset_t \subseteq N_{t-1}$ and $\Aset_t \subseteq N_{t-1}^c$, (b) $N_t = N_{t-1} \cup \Aset_t \setminus \Rset_t$, and (c) if two sets $B,C$ are disjoint, then, $D \cup C \setminus B :=(D \cup C) \setminus B =(D \cap B^c) \cup C$ for any set $D$.%

We first bound $|T_t|$, $|\Delta_{e,t}|$, $|\Delta_t|$. Since $T_t = \tT_{t-1} = \Nhat_{t-1}$, so $|T_t| \le S_0$. Also, $\Delta_{e,t} = \Nhat_{t-1} \setminus N_t =  \Nhat_{t-1} \cap [(N_{t-1}^c \cap \Aset_t^c) \cup \Rset_t] \subseteq \tDelta_{e,t-1} \cup \Rset_t = \Rset_t$. The last equality follows since $|\tDelta_{e,t-1}| =0$. Thus $|\Delta_{e,t}| \le |\Rset_t| = S_a$.

Consider $|\Delta_t|$. Notice that $\Delta_t = N_t \setminus \Nhat_{t-1}  = (N_{t-1} \cap \Nhat_{t-1}^c \cap \Rset_t^c) \cup (\Aset_t \cap \Nhat_{t-1}^c) = (\tDelta_{t-1} \cap  \Rset_t^c) \cup ( \Aset_t \cap \Nhat_{t-1}^c) \subseteq  (\Sset_{t-1}(2)  \cap  \Rset_t^c) \cup  \Aset_t = \Sset_{t-1}(2) \cup  \Aset_t \setminus \Rset_t$. Here we used $\tDelta_{t-1} \subseteq \Sset_{t-1}(2)$.
 Since $\Rset_t \subseteq \Sset_{t-1}(2)$ and $\Aset_t$ is disjoint with $\Sset_{t-1}(2)$, thus $|\Delta_t| \le |\Sset_{t-1}(2)| + |\Aset_t| - |\Rset_t| = 2S_a + S_a - S_a$.

Next we bound $|\tDelta_t|$, $|\tDelta_{e,t}|$, $|\tT_t|$. Consider the support estimation step. Apply the first claim of Lemma \ref{lemma_modcs} with $S_N=S_0$, $S_{\Delta e}=S_a$, $S_\Delta = 2S_a$, and $b_1 = 2r$. Since conditions \ref{measmodel_simple} and \ref{add_del_simple} of the theorem hold, all elements of $N_t$ with magnitude equal to or greater than $2r$ will get detected. Thus, $\tDelta_t \subseteq \Sset_t(2)$. Apply the second claim of the lemma. Since conditions \ref{measmodel_simple} and \ref{threshes_simple} hold, all zero elements will get deleted and there will be no false detections, i.e. $|\tDelta_{e,t}|=0$. Finally, $|\tT_t| \le |N_t| + |\tDelta_{e,t}| \le S_0 + 0$.

The second claim for time $t$ follows using the first claim for time $t-1$ and the arguments from the paras above. The third claim follows using the second claim and Corollary \ref{modcs_cs_bnd}. 

\subsection{Appendix: Proof of Theorem \ref{stabres_modcs}}
\label{proof_addLSdel_modcs}

We prove the first claim of the theorem by induction. Using condition \ref{initass} of the theorem, the claim holds for $t=0$. This proves the base case. For the induction step, assume that the claim holds at $t-1$, i.e. $|\tDelta_{e,t-1}| =0$, $|T_{t-1}| \le S_0$, and $\tDelta_{t-1} \subseteq \Sset_{t-1}(2)$ so that $|\tDelta_{t-1}| \le 2S_a$. Using this, we prove that the claim holds at $t$. We will use the following facts often: (a) $\Rset_t \subseteq N_{t-1}$,  (b) $\Aset_t \subseteq N_{t-1}^c$, (c) $N_t = N_{t-1} \cup \Aset_t \setminus \Rset_t$, and (d) if two sets $B,C$ are disjoint, then, $D \cup C \setminus B :=(D \cup C) \setminus B =(D \cap B^c) \cup C$ for any set $D$.%

The bounding of $|T_t|, |\Delta_t|, |\Delta_{e,t}|$ is exactly as in the proof of Theorem \ref{stabres_simple_modcs}. Since $T_t = \tT_{t-1}$, so $|T_t| \le S_0$. Also, $\Delta_{e,t} = \Nhat_{t-1} \setminus N_t =  \Nhat_{t-1} \cap [(N_{t-1}^c \cap \Aset_t^c) \cup \Rset_t] \subseteq \tDelta_{e,t-1} \cup \Rset_t = \Rset_t$. 
Thus $|\Delta_{e,t}| \le |\Rset_t| = S_a$.
Finally, $\Delta_t = N_t \setminus \Nhat_{t-1}  =  (\tDelta_{t-1} \cap  \Rset_t^c) \cup ( \Aset_t \cap \Nhat_{t-1}^c) \subseteq  (\Sset_{t-1}(2)  \cap  \Rset_t^c) \cup  \Aset_t$. Thus,
\bea
\Delta_t \subseteq \Sset_{t-1}(2) \cup  \Aset_t \setminus \Rset_t
\label{Delta_sub_0}
\eea
Since $\Rset_t \subseteq \Sset_{t-1}(2)$ and $\Aset_t$ is disjoint with $\Sset_{t-1}(2)$, thus $|\Delta_t| \le |\Sset_{t-1}(2)| + |\Aset_t| - |\Rset_t| = 2S_a + S_a - S_a$.

Consider the detection step. There are at most $S_a$ false detects (from condition \ref{addthresh}) and thus $|\tDelta_{e,\dett,t}| \le |\Delta_{e,t}| + S_a \le 2S_a$. Thus $|T_{\dett,t}| \le |N_t| + |\tDelta_{e,\dett,t}| \le S_0+2S_a$.

Next, consider $|\Delta_{\dett,t}|$. Notice that
\bea
\Delta_t \subseteq \Sset_{t-1}(2) \cup \Aset_t \setminus \Rset_t \subseteq \Sset_t(2) \cup \Iset_t(2) \setminus \Dset_t(1).
\label{Delta_sub}
\eea
The first $\subseteq$ is from (\ref{Delta_sub_0}), the second one follows by using (\ref{sseteq_2}) for $j=2$. Now, apply Lemma \ref{detectcond_modcs} with $S_N = S_0$, $S_{\Delta e}=S_a$, $S_\Delta = 2S_a$, and with $b_1 = 2r$. Using (\ref{Delta_sub}), $L = \Delta_t \cap \Iset_t(2)$.
Since conditions \ref{measmodel} and \ref{add_del} hold, by Lemma \ref{detectcond_modcs}, all elements of $L$ will definitely get detected at time $t$. Thus $\Delta_{\dett,t} \subseteq \Delta_t \setminus L =  \Delta_t \setminus \Iset_t(2)$. But from (\ref{Delta_sub}), $\Delta_t \setminus \Iset_t(2) \subseteq \Sset_t(2) \setminus \Dset_{t}(1)$.
Since $\Dset_{t}(1) \subseteq \Sset_{t}(2)$, so $|\Delta_{\dett,t}| \le |\Sset_t(2)| - |\Dset_{t}(1)|= 2S_a - S_a$.

Consider the deletion step. Apply  Lemma \ref{truedelscond} with $S_T = S_0+2S_a$, $S_\Delta = S_a$. Since condition \ref{measmod_delta} holds, $\delta_{S_0+2S_a} < 1/2$ holds. Since $\Delta_{\dett,t} \subseteq \Sset_t(2) \setminus \Dset_{t}(1)$, $\Delta_{\dett,t}$ contains at most $S_a$ elements of magnitude $r$ and nothing else. Thus,  $\|(x_t)_{\Delta_{\dett,t}}\| \le \sqrt{S_a}r$. Using these facts and condition \ref{delthresh}, by Lemma \ref{truedelscond}, all elements of $\tDelta_{e,\dett,t}$ will get deleted. Thus $|\tDelta_{e,t}|=0$. Thus $|\tT_t| \le |N_t| + |\tDelta_{e,t}| \le S_0$.%

To bound $|\tDelta_{t}|$, apply Lemma \ref{nofalsedelscond} with $S_T = S_0+2S_a$, $S_\Delta = S_a$, $b_1 = 2r$. 
By Lemma \ref{nofalsedelscond}, to ensure that all elements of $L$ do not get falsely deleted, we need $\delta_{S_0+2S_a} < 1/2$ and $2r > \alpha_{\del} + \sqrt{2} \eps + 2 \theta_{S_0+2S_a,S_a} \sqrt{S_a} r$. From condition \ref{delthresh}, $\alpha_{\del} =  \sqrt{2} \eps + 2 \theta_{S_0+2S_a,S_a} \sqrt{S_a} r$. Thus, we need $\delta_{S_0+2S_a} < 1/2$ and $2r > 2(\sqrt{2} \eps + 2 \theta_{S_0+2S_a,S_a} \sqrt{S_a} r)$. $\delta_{S_0+2S_a} < 1/2$ holds since condition \ref{measmod_delta} holds. The second condition holds since condition \ref{measmod_theta} and $r \ge G_2$ of condition \ref{add_del} hold. Thus, we can ensure that all elements of $L$, i.e. all elements of $T_{\dett,t}$ with magnitude greater than or equal to $b_1=2r$ do not get falsely deleted.
But nothing can be said about the elements smaller than $2r$ (in the worst case all of them may get falsely deleted). Thus, $\tDelta_t \subseteq \Sset_t(2)$ and so $|\tDelta_t| \le 2S_a$.

 This finishes the proof of the first claim. To prove the second and third claims for any $t>0$: use the first claim  for $t-1$ and the arguments from the paragraphs above to show that the second and third claim hold for $t$.
The fourth claim follows directly from the first claim and fact \ref{errls1} of Proposition \ref{prop1} (applied with  $x \equiv \xhat_t$, $T \equiv \tT_t$, $\Delta \equiv \tDelta_t$). The fifth claim follows directly from the second claim and Corollary \ref{modcs_cs_bnd}.



\subsection{Appendix: Generalized version of Corollary \ref{cor2_relax}}
\label{stabres_modcs_gen}

\begin{corollary}[Stability of modified-CS with add-LS-del -- 3] 
Assume Signal Model \ref{sigmod2} and  $\|w_t\| \le \eps$. Let $e_t:=(x_t - \xhat_{\dett,t})_{T_{\dett,t}}$. Assume that the LS step error is spread out enough so that $$\|e_t\|_\infty \le \frac{\zeta_m}{\sqrt{S_a}} \|e_t\|$$ at all $t$.
Consider Algorithm \ref{modcsalgo_2}.
If, for some $1 \le d_0 \le d$, 
\ben
\item {\em (addition and deletion thresholds) }
\ben
\item $\alpha_{\dett}$ is large enough so that there are at most $f$ false additions per unit time,
\label{addthresh}

\item $\alpha_{\del}  = \sqrt{\frac{2}{S_a}}\zeta_m \eps +  2  k_3  \theta_{S_0+S_a+f,k_2}\zeta_m r  $,
\label{delthresh}
\een

\item {\em (support size, support change rate)} $S_0,S_a$ satisfy
\ben
\item $\delta_{S_0 + S_a(1 + k_1) } < (\sqrt{2}-1)/2$ and $S_a \le \frac{S_0}{3k_1}$,
\item $\delta_{S_0+S_a + f} < 1/2$,
\item $\theta_{S_0+S_a+f,k_2S_a} < \frac{1}{2} \frac{d_0}{4k_3 \zeta_m}$,
\label{theta_ass_0}
\een
\label{measmodel}

\item {\em (new element increase rate) } $r \ge \max(\check{G}_1,\check{G}_2)$, where
\label{add_del}
\bea
\check{G}_1 \sdefn \frac{ \alpha_{\dett} + 8.79\eps }{d_0}  \nn \\ 
\check{G}_2 \sdefn \frac{2\sqrt{2} \zeta_m \eps}{\sqrt{S_a} (d_0 -  4k_3 \theta_{S_0+S_a+f,k_2S_a}\zeta_m) }  \ \ \ \ \ \ \
\eea
\item {\em (initial time)} $n_0$ is large enough to ensure that $\tDelta_0  \subseteq \Sset_0(d_0)$, $|\tDelta_0| \le (2d_0-2)S_a$,  $|\tDelta_{e,0}| =0$, $|\tT_0| \le S_0$,
\label{initass}
\een
where
\bea
k_1 \sdefn \max(1,2d_0-2) \nn \\
k_2 \sdefn \max(0,2d_0-3) \nn \\
k_3 \sdefn \sqrt{ \sum_{j=1}^{d_0-1} j^2 +  \sum_{j=1}^{d_0-2} j^2 }  
\eea
then,
\ben
\item  at all $t \ge 0$, $|\tT_t| \le S_0$, $|\tDelta_{e,t}| =0$, and $\tDelta_t \subseteq \Sset_t(d_0)$ and so $|\tDelta_t| \le (2d_0-2)S_a$,

\item at all $t > 0$, $|T_t| \le S_0$, $|\Delta_{e,t}| \le S_a$, and $|\Delta_t| \le k_1 S_a$,

\item at all $t > 0$, $|T_{\dett,t}| \le S_0+S_a+f$, $|\Delta_{e,\dett,t}| \le S_a+f$, and $|\tDelta_{\dett,t}| \le k_2 S_a$

\item at all $t > 0$, $\|x_t-\xhat_t\| \le \sqrt{2} \eps +  k_3  \sqrt{S_a} (2\theta_{S_0,(2d_0-2)S_a}+1) r  $

\item  at all $t > 0$, $\|x_t - \xhat_{t,modcs}\| \le C_1(S_0+S_a + k_1 S_a) \eps \le 8.79 \eps$.
\een
\label{gencase}
\end{corollary}

{\em Proof: } The proof follows using exactly the same steps as in the proof of Theorem \ref{stabres_modcs}, but of course with Lemmas \ref{nofalsedelscond} and \ref{truedelscond} replaced by Lemmas \ref{nofalsedels_equalLS} and \ref{truedelscond_equalLS} respectively.
The only difference is that, instead of ensuring $|\tDelta_{e,t}|=0$ and $\tDelta_t \subseteq \Sset_t(2)$, we try to ensure $|\tDelta_{e,t}|=0$ and $\tDelta_t \subseteq \Sset_t(d_0)$ for some $d_0 \le d$. For $1 < d_0 \le d$, notice that $|\Sset_{t}(d_0)| = (2d_0-2) S_a$. Also, since, now, $\Delta_{\dett,t} \subseteq  \Sset_{t}(d_0) \setminus \Dset_t(d_0-1)$, so $|\Delta_{\dett,t}| \le (2d_0-3)S_a$ and $\|x_{\dett,t}\| \le k_3 S_a$.
The case of $d_0=1$ is handled separately. In this case, $\Sset_{t}(d_0)$ is empty, but still $\Delta_t$ is not empty, but is equal to $\Aset_t$. Also, $\Delta_{\dett,t}$ and $\tDelta_t$ are empty.


\subsection{Proof of Lemma \ref{detectcond_lscs}}
\label{proof_detectcond_lscs}
From Lemma \ref{lscs_bnd}, if $\|w\| \le \eps$, $\delta_{2|\Delta|} < (\sqrt{2}-1)/2$ and $\delta_{|T|} < 1/2$, then $\|x - \xhat_{\CSres}\| \le C'(|T|,|\Delta|) \eps +  \theta_{|T|,|\Delta|} C''(|T|,|\Delta|) \|x_\Delta\|$.
Using the fact that $\|x_\Delta\| \le \sqrt{|L_1|} b + \|x_{\Delta_2}\|$; fact \ref{det1} of Proposition \ref{prop1}; and the fact that for all $i \in L_1$, $|x_i| \ge \gamma b$, we can conclude that all $i \in L_1$ will get detected if
\ben
\item $\delta_{2|\Delta|} < (\sqrt{2}-1)/2$,
\item $\delta_{|T|} < 1/2$ and
\item $\alpha_\dett + C' \eps + \theta C'' (\sqrt{|L_1|} b + \|x_{\Delta_2}\|) < \gamma b$. Using $\|x_{\Delta_2}\| \le \kappa b$ and $|L_1| \le S_{L 1}$, this inequality holds if
\ben
\item $\theta C'' \le \frac{\gamma}{2(\sqrt{S_{L 1}} + \kappa)}$ and
\item $\frac{ \alpha_\dett + C'\eps }{\gamma - \theta C'' (\sqrt{S_{L 1}} + \kappa)} < b$.
\een
\een
Since we only know that $|T| \le S_T$, $|\Delta| \le S_\Delta$, we need the above inequalities to hold for all values of $|T|,|\Delta|$ satisfying these upper bounds. This leads to the conclusion of the lemma. Notice that the LHS's the first two inequalities are non-decreasing functions of  $|\Delta|,|T|$ and thus the lemma just uses their upper bounds. The LHS's of the last two are non-decreasing in $|T|$, but are not monotonic in $|\Delta|$ (since $C'(|T|,|\Delta|)$ and $C''(|T|,|\Delta|)$ are not monotonic in $|\Delta|$). Hence we explicitly maximize over $|\Delta| \le S_\Delta$.

\subsection{Proof of Lemma \ref{modcsbnd}}
\label{modcserrorbound}
We provide the proof here for the sake of completion and for ease of review. This will be removed later.
Let $h : = \xhat_{modcs} - x$. We adapt the approach of \cite{candes_rip} to bound the reconstruction error, $\|h\| := \|\xhat-x\|$. A similar result was obtained in \cite{arxiv}.
Let $\Delta_1$ denote the set of  indices of $h$ with the $|\Delta|$ largest values outside of $T \cup \Delta$, let $\Delta_2$ denote the indices of the next $|\Delta|$ largest values and so on. Then using the same approach as that of \cite{candes_rip},
\bea
\|h_{(T \cup \Delta \cup \Delta_1)^c}\| \sle \sum_{j \ge 2} \|h_{\Delta_j}\|
                                          \le \frac{1}{\sqrt{|\Delta|}} \|h_{(T \cup \Delta)^c}\|_1
\label{h1}
\eea
Since $\xhat_{modcs} = x+h$ is the minimizer of (\ref{modcs}) and since both $x$ and $\xhat_{modcs}$ are feasible; and since $x$ is supported on $N \subseteq T \cup \Delta$,
\bea
\|x_\Delta\|_1=\|x_{T^c}\|_1 \sge  \|(x+h)_{T^c}\|_1 \nn \\
\sge \|x_\Delta\|_1 - \|h_\Delta\|_1 + \|h_{(T \cup \Delta)^c}\|_1  
\eea
Thus,
\bea
\|h_{(T \cup \Delta)^c}\|_1 \le \|h_\Delta\|_1    
\eea
Combining this with (\ref{h1}), and using $\frac{\|h_\Delta\|_1}{\sqrt{|\Delta|}} \le \|h_\Delta\|$, we get
\bea
\|h_{(T \cup \Delta \cup \Delta_1)^c}\| \sle \sum_{j \ge 2} \|h_{\Delta_j}\| \le \|h_\Delta\|
\label{h2}
\eea

Next, since both $x$ and $\xhat_{modcs}$ are feasible,
\bea
\|Ah\| \se \|A(x - \xhat_{modcs})\| \nn \\
\sle \|y - Ax\| + \|y - A \xhat_{modcs}\| \le 2 \eps
\label{feas}
\eea
In this proof, let
\bea
\delta \sdefn \delta_{|T| + 2|\Delta|} \quad \text{~and~} \quad
\theta \defn \theta_{|T|,|\Delta|}
\eea

Now, we upper bound $\|h_{T \cup \Delta \cup \Delta_1}\|$. To do that, notice that
\bea
(1-\delta) \|h_{T \cup \Delta \cup \Delta_1}\|^2 \le  \|Ah_{T \cup \Delta \cup \Delta_1}\|^2
\eea
To bound the RHS of the above, notice that $Ah_{T \cup \Delta \cup \Delta_1} = Ah - \sum_{j \ge 2} A h_{\Delta_j}$ and so
\bea
&& \|Ah_{T \cup \Delta \cup \Delta_1}\|^2  = \langle Ah_{T \cup \Delta \cup \Delta_1}, Ah \rangle - \sum_{j \ge 2}\langle A h_{T \cup \Delta \cup \Delta_1}, A h_{\Delta_j} \rangle \nn
\eea
Using (\ref{feas}) and the definition of $\delta_S$ given in (\ref{def_delta}),
\bea
|\langle A h_{T \cup \Delta \cup \Delta_1}, Ah \rangle | \sle 2\eps \sqrt{1+\delta}\|h_{T \cup \Delta \cup \Delta_1}\|
\eea
Using the definition of $\theta_{S_1,S_2}$ given in (\ref{def_theta}); equation (\ref{h2}); and the fact that $\|h_{T}\| + \|h_{\Delta \cup \Delta_1}\| \le \sqrt{2}\|h_{T \cup \Delta \cup \Delta_1}\|$, we get the following. If $2|\Delta| \le |T|$,
\bea
&& |\sum_{j \ge 2} \langle A h_{T \cup \Delta \cup \Delta_1}, A h_{\Delta_j} \rangle | \nn \\
&&  \le (\theta \|h_{T}\| + \theta_{2|\Delta|,\Delta|}\|h_{\Delta \cup \Delta_1}\|) \sum_{j \ge 2} \|h_{\Delta_j}\| \nn \\
&& \le  \sqrt{2} \theta \|h_{T \cup \Delta \cup \Delta_1}\| \ \|h_\Delta\|    
\label{b1}
\eea
Combining the last four equations above, if $2|\Delta| \le |T|$,
\bea
(1-\delta) \|h_{T \cup \Delta \cup \Delta_1}\| \le 2\eps \sqrt{1+\delta} +  \sqrt{2} \theta  \|h_\Delta\|
\eea
Using $\|h_\Delta\| \le \|h_{T \cup \Delta \cup \Delta_1}\|$, we can simplify the above to get
\bea
\|h_{T \cup \Delta \cup \Delta_1}\| \le \frac{2 \sqrt{1+\delta}}{1-\delta - \sqrt{2} \theta} \eps
\eea
Finally, using (\ref{h2}) and $\|h_\Delta\| \le \|h_{T \cup \Delta \cup \Delta_1}\|$ and the above, 
\bea
\|h\| 
\sle 2\|h_{T \cup \Delta \cup \Delta_1}\|  \le \frac{4 \sqrt{1+\delta}}{1-\delta - \sqrt{2} \theta} \eps 
\eea

Clearly, all of the above discussion holds only if the RHS is positive which is true only if $\delta + \sqrt{2} \theta < 1$. Also, (\ref{b1}) and hence everything after that needs $2|\Delta| \le |T|$. Since $|T| = |N|+|\Delta_e| - |\Delta|$, this will hold if $3|\Delta| \le |N|$. Thus, we get the following result.
\begin{corollary}
If  $|\Delta| \le |N|/3$ and if $\delta_{|T|+2|\Delta|} + \sqrt{2} \theta_{|T|,|\Delta|} < 1$, then
\bea
\|h\| \sle  \frac{4 \sqrt{1+\delta}}{1-\delta_{|T|+2|\Delta|} - \sqrt{2} \theta_{|T|,|\Delta|}} \eps
\eea
\label{corlemma1}
\end{corollary}
Using $\theta_{|T|,|\Delta|} \le \delta_{|T|+|\Delta|} \le \delta_{|T|+2|\Delta|}$ \cite{decodinglp} in both the required sufficient condition and in the bound; and by substituting $|T| = |N|+|\Delta_e| - |\Delta|$; and by using $\frac{1}{\sqrt{2}+1} =\sqrt{2}-1$ we get the notationally simpler result of Lemma \ref{modcsbnd}.

%

\begin{figure*}[h!]
\centerline{
\subfigure[$n=65$, $r=1$, $d=3$]{
\label{simfig_n65_r1}
\begin{tabular}{ccc}
\epsfig{file = 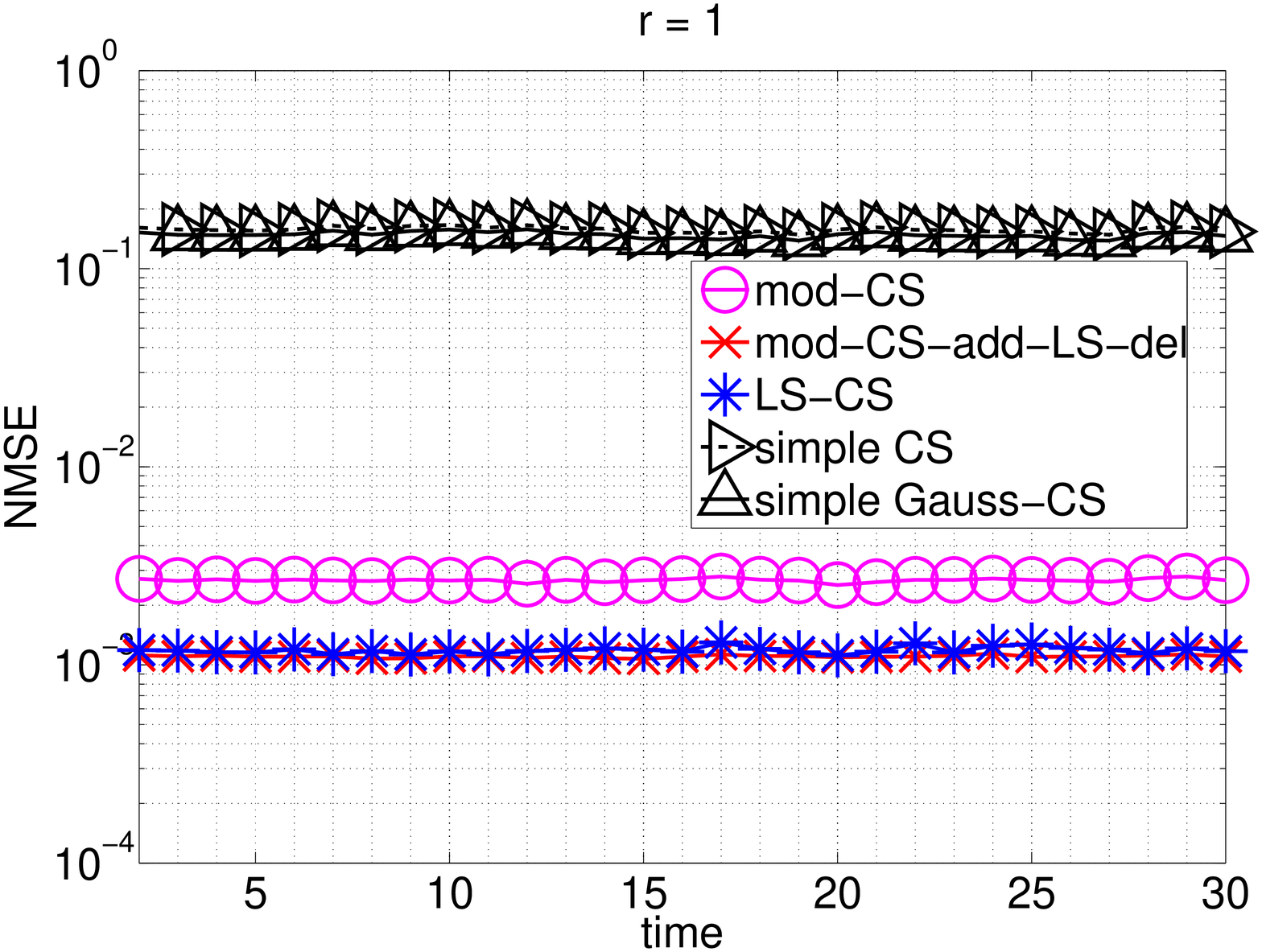, width=5.5cm } &
\epsfig{file = 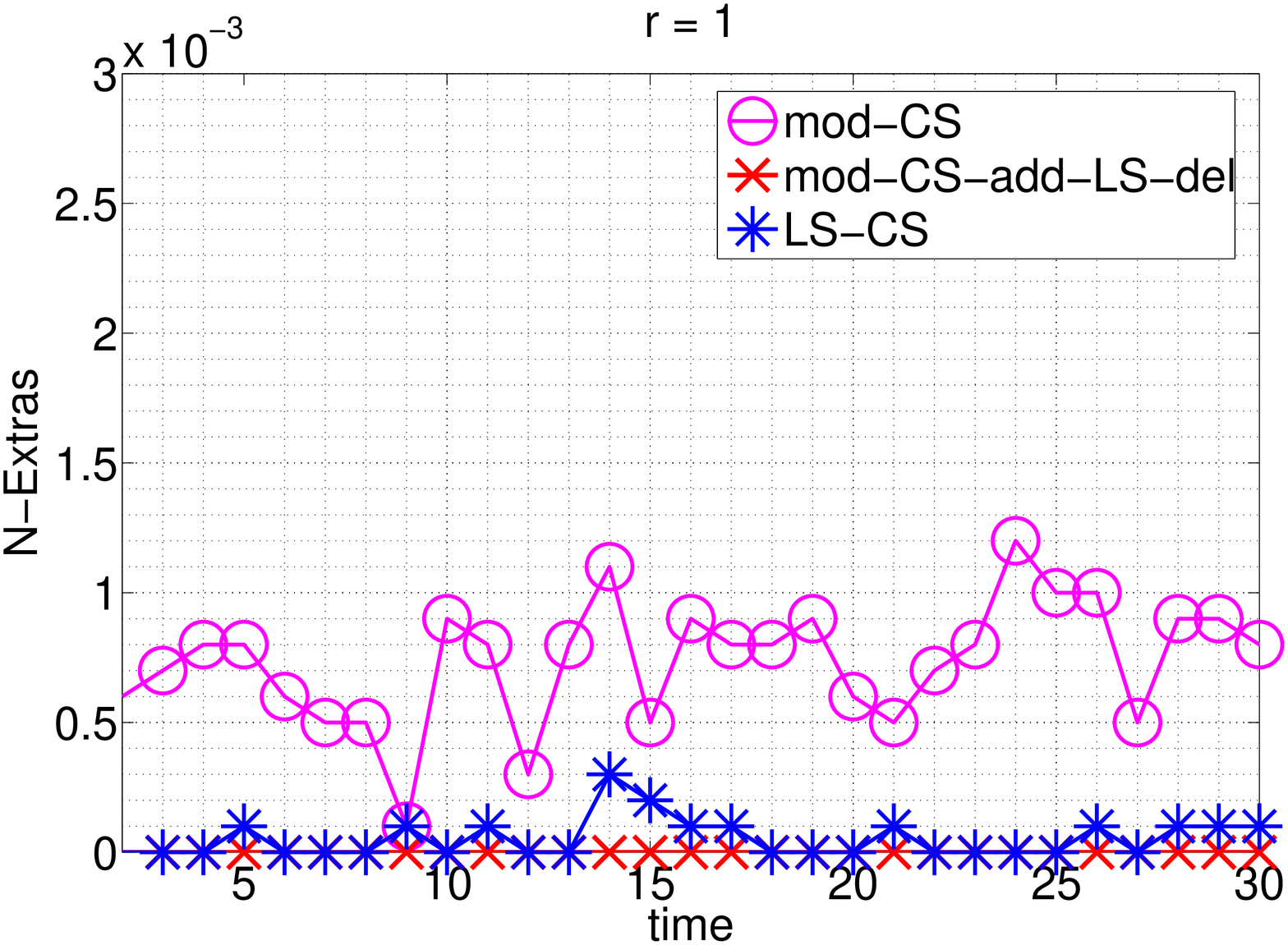, width=5.5cm} &
\epsfig{file = 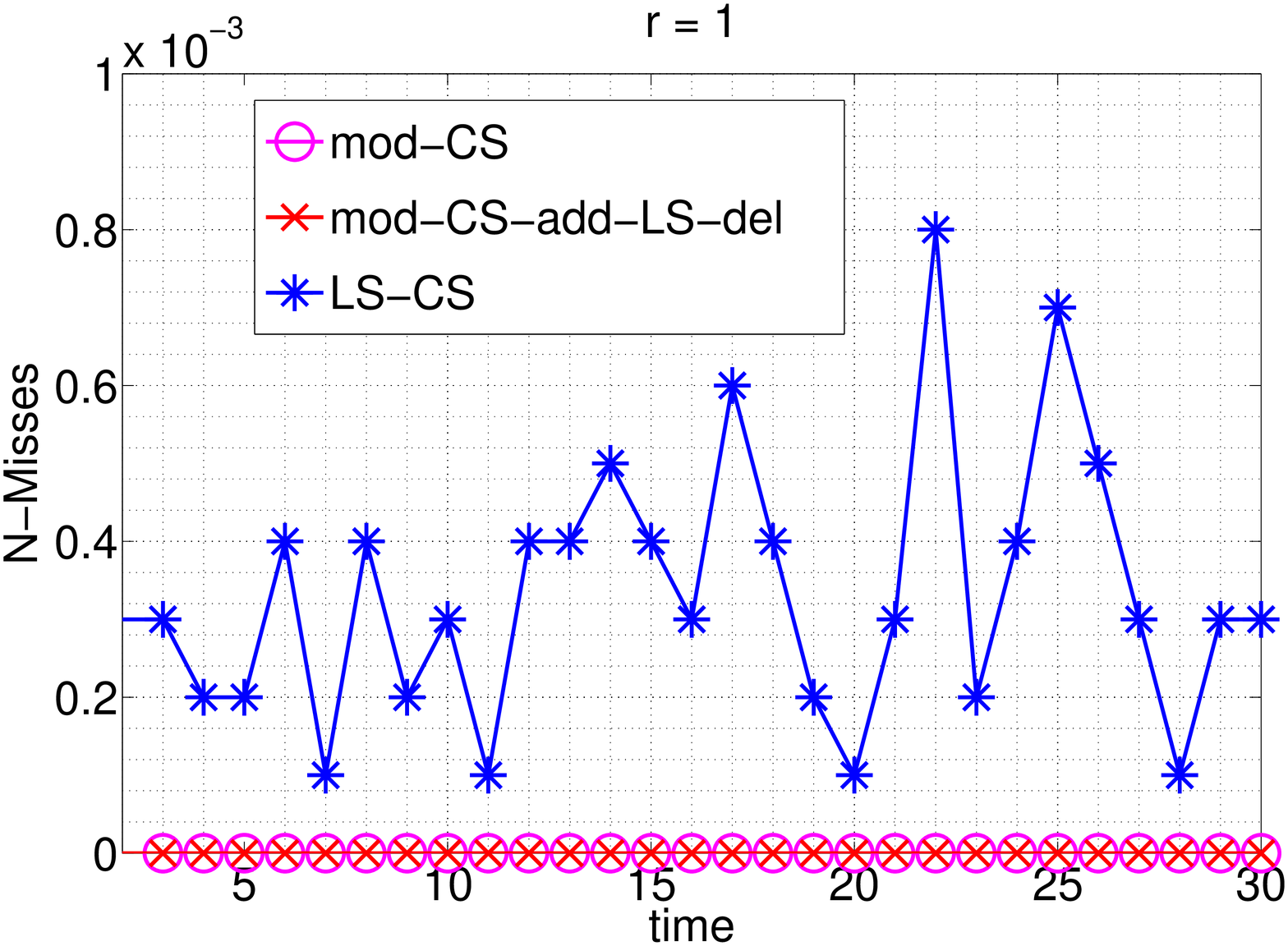, width=5.5cm}
\end{tabular}
}
}
\centerline{
\subfigure[$n=59$, $r=1$, $d=3$]{
\label{simfig_n59_r1}
\begin{tabular}{ccc}
\epsfig{file = 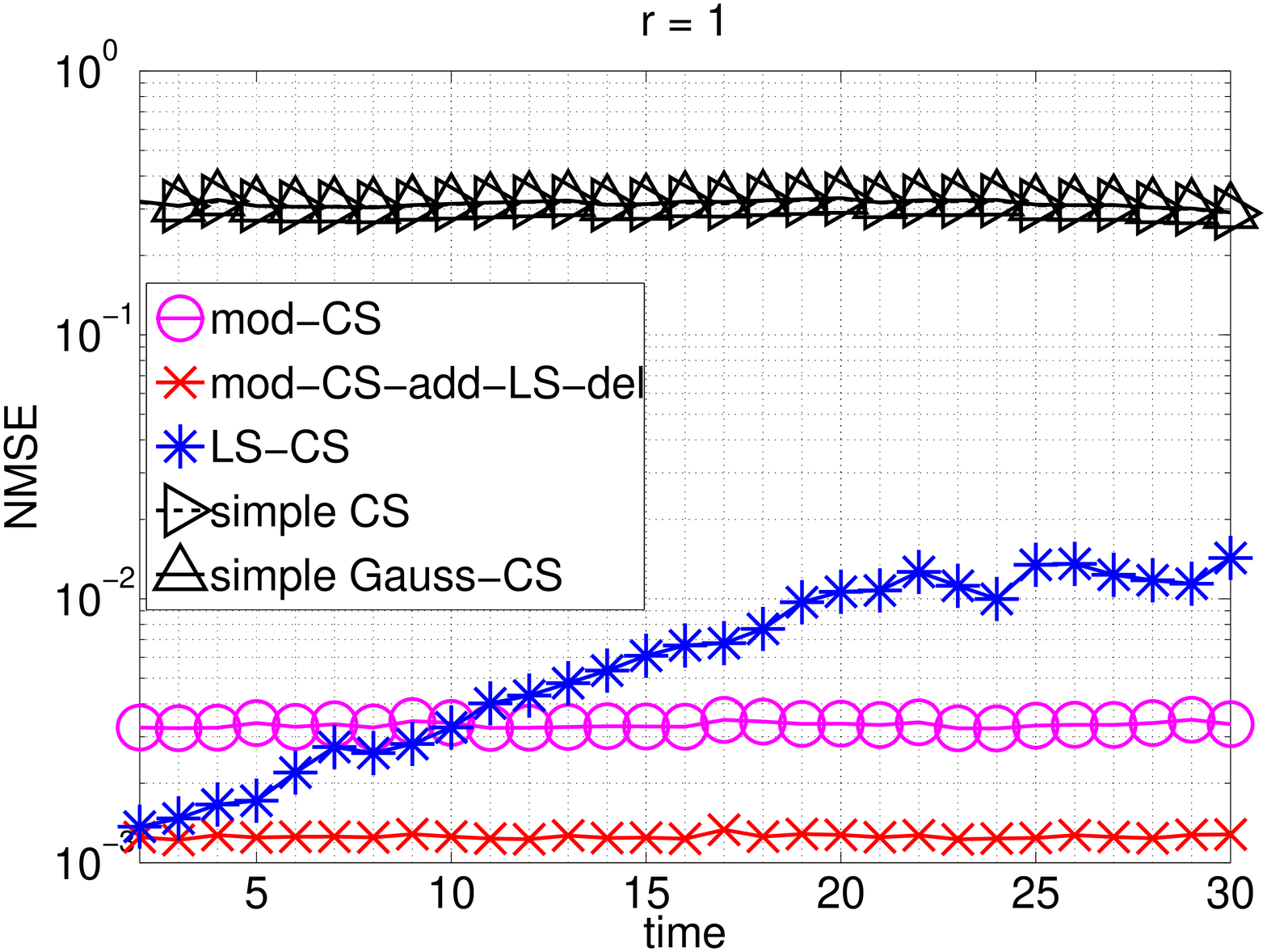, width=5.5cm } &
\epsfig{file = 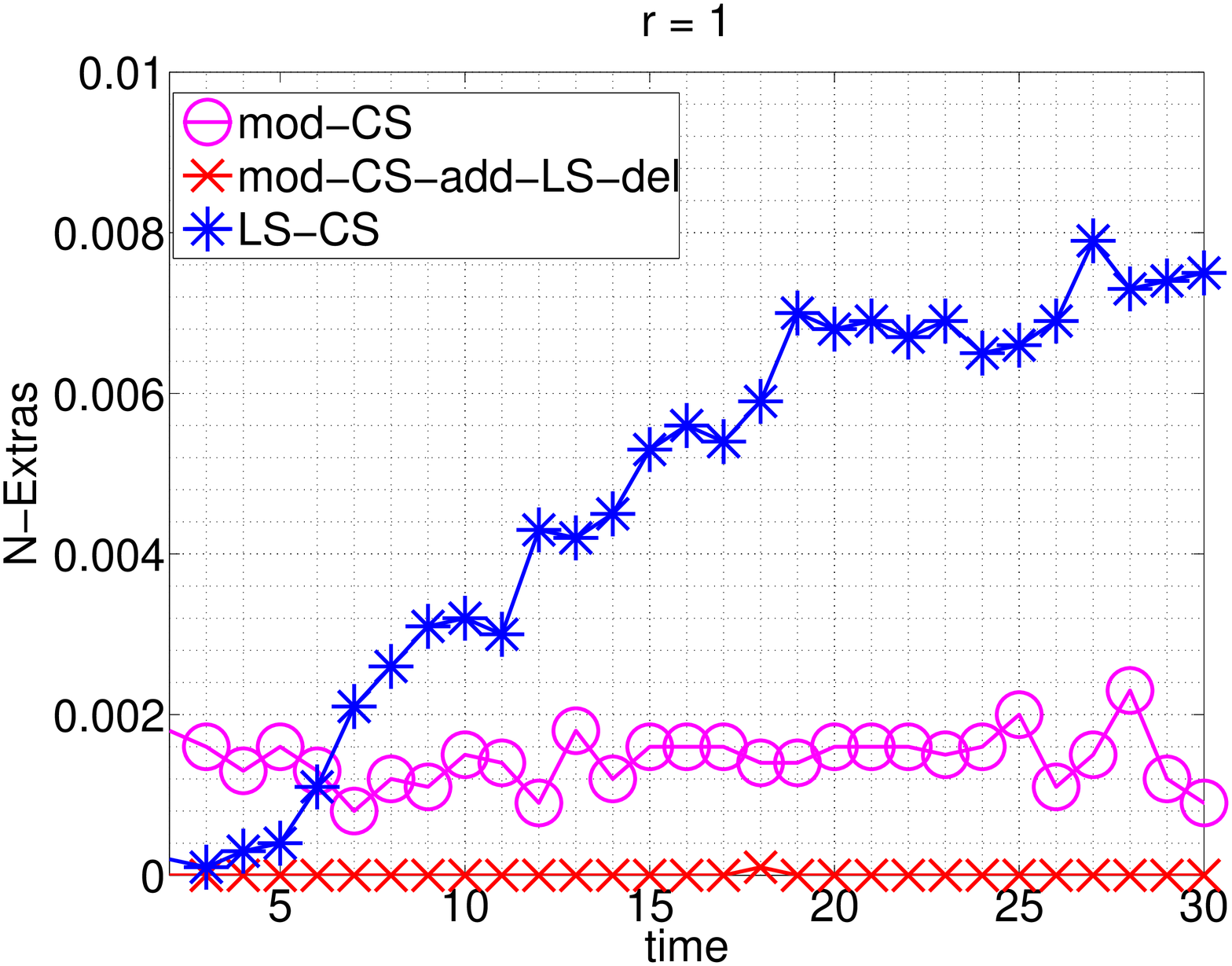, width=5.5cm} &
\epsfig{file = 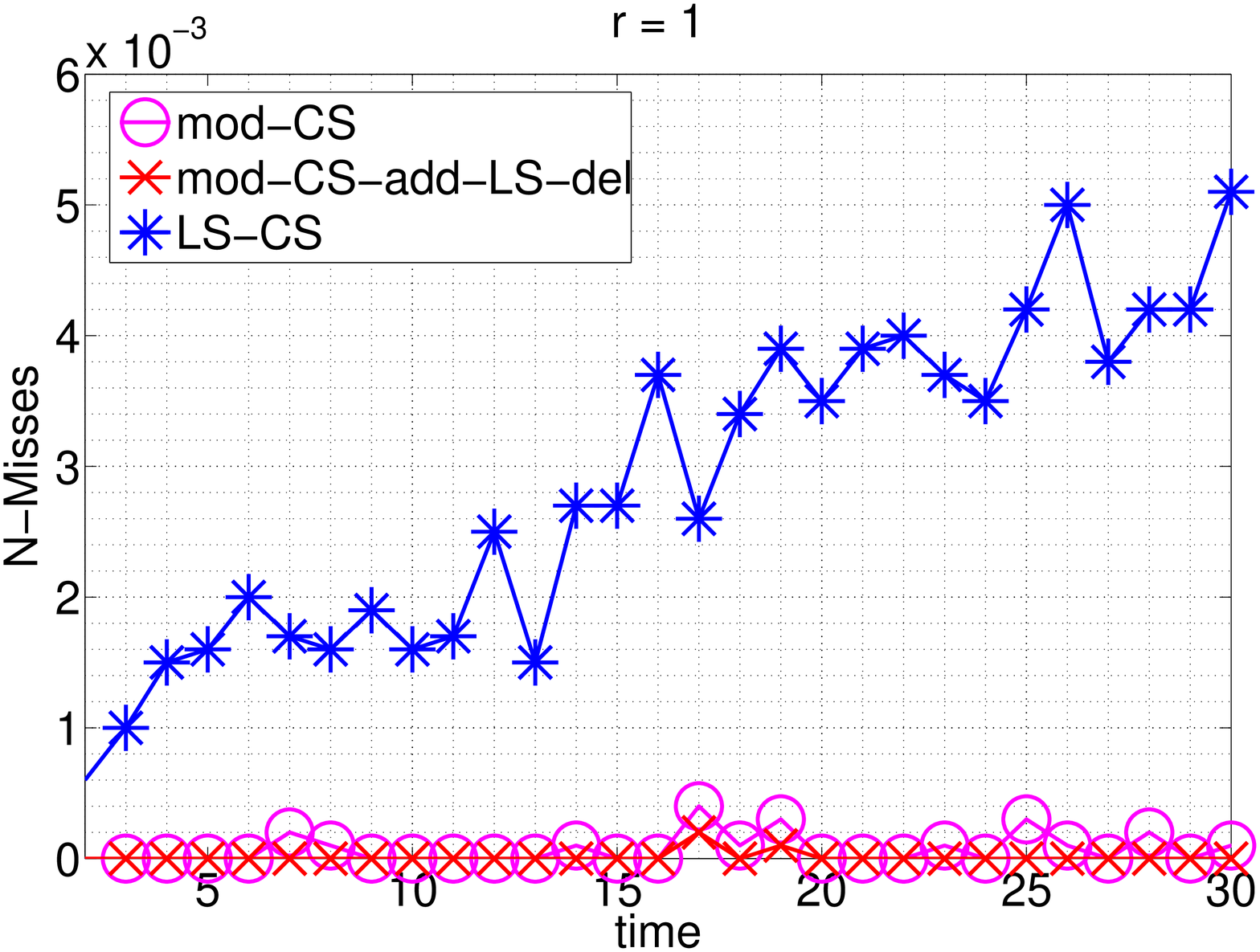, width=5.5cm}
\end{tabular}
}
}
\centerline{
\subfigure[$n=59$, $r=2/3$,  $d=3$]{
\label{simfig_n59_r2b3}
\begin{tabular}{ccc}
\epsfig{file = 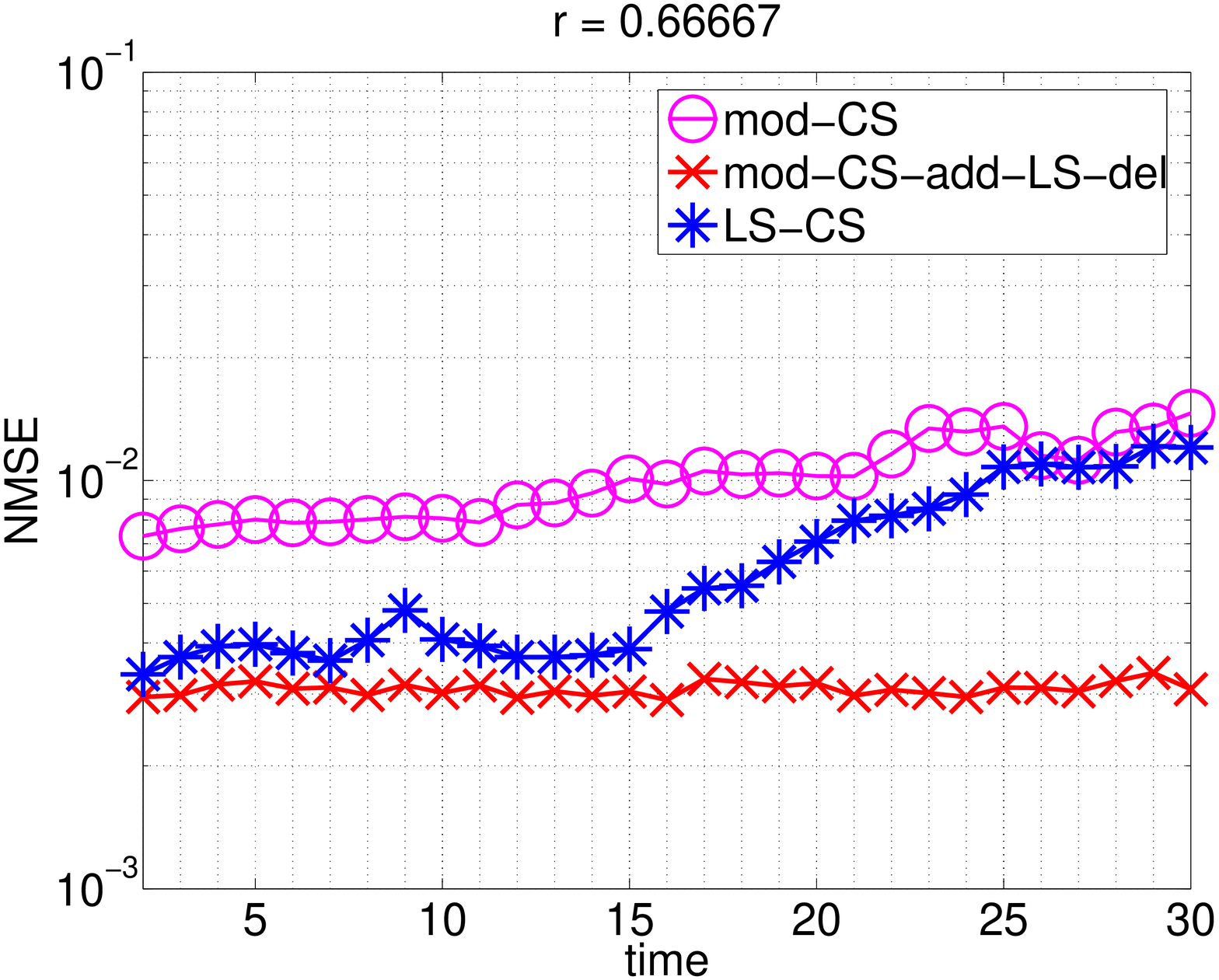, width=5.5cm} &
\epsfig{file = 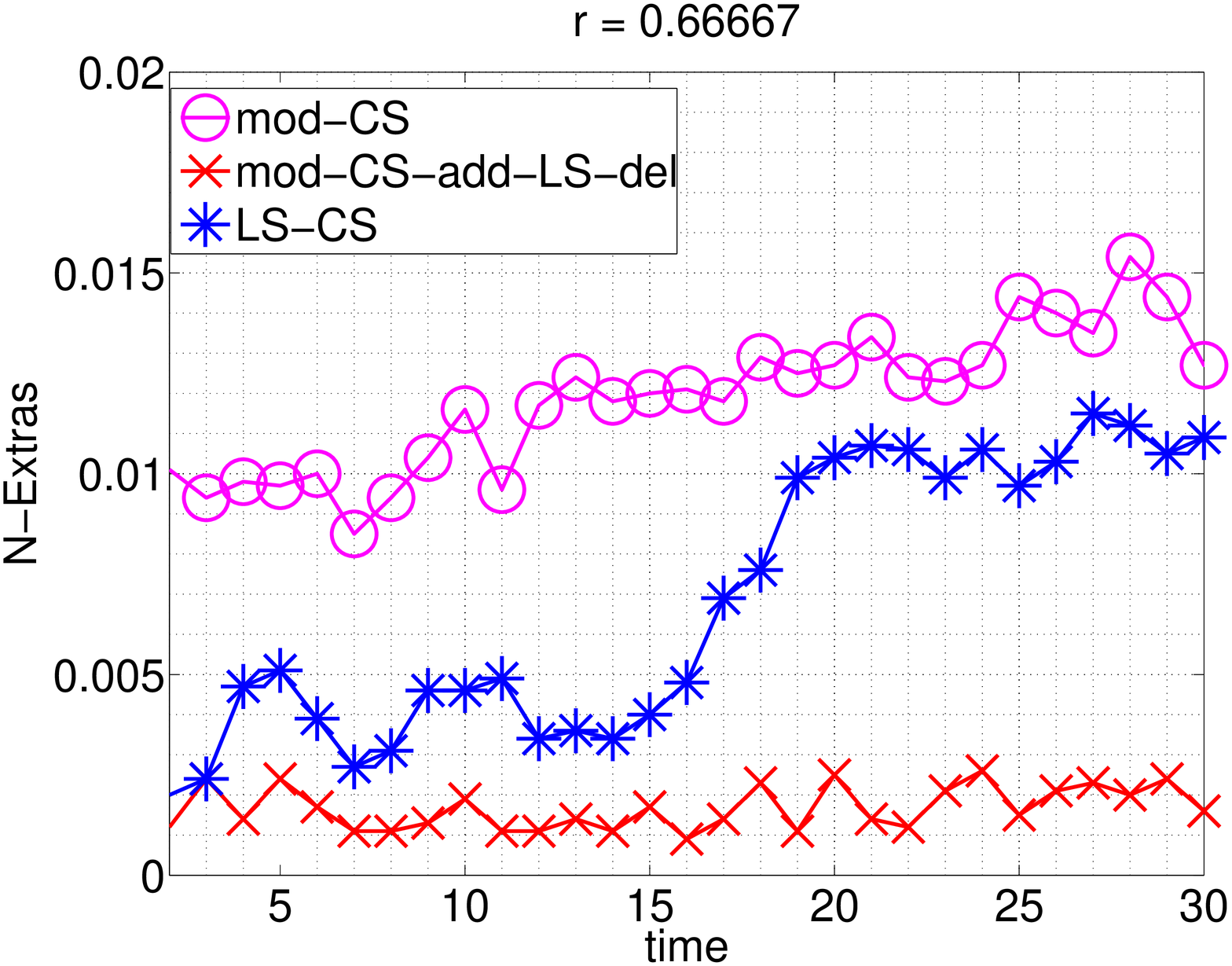, width=5.5cm} &
\epsfig{file = 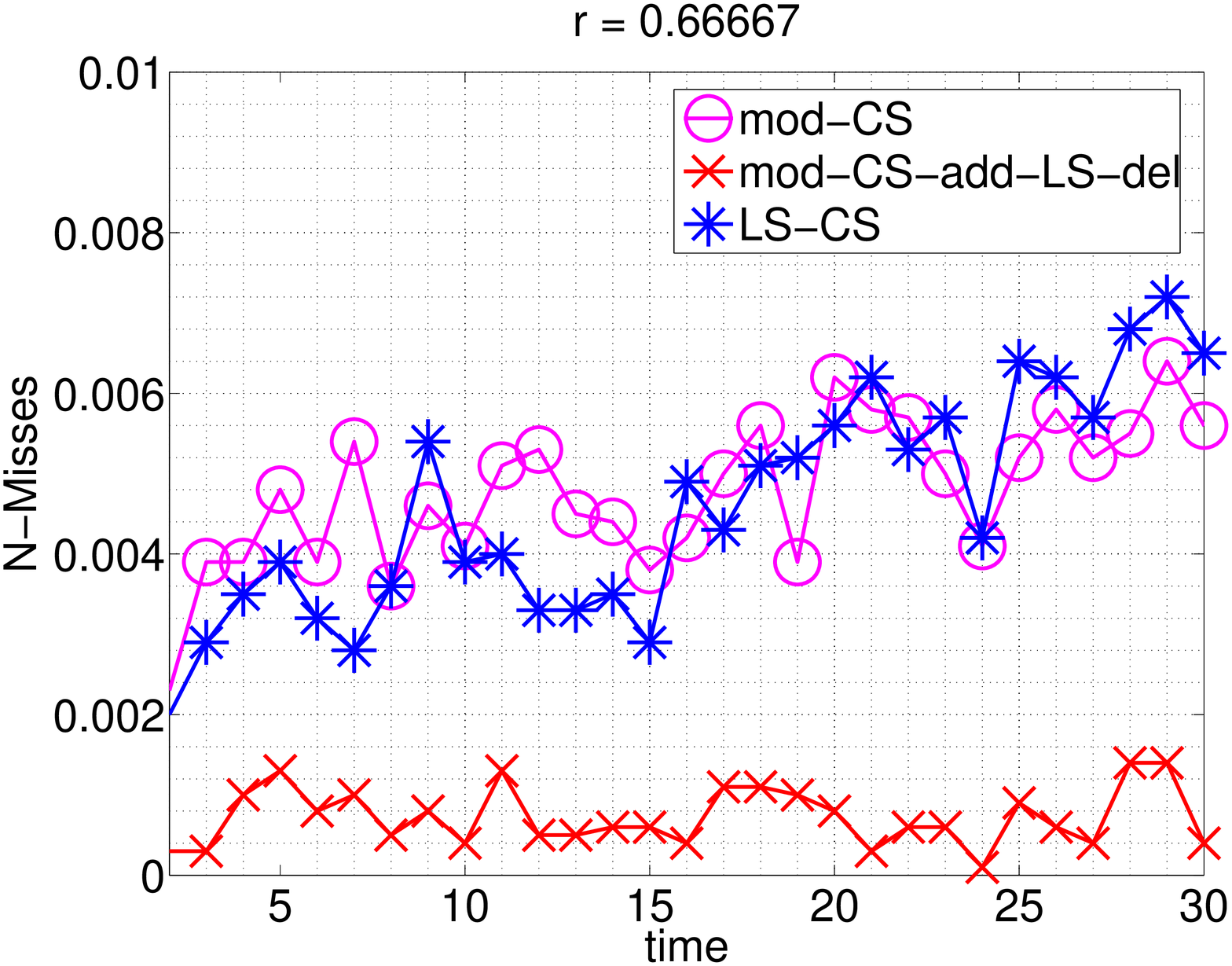, width=5.5cm}
\end{tabular}
}
}
\centerline{
\subfigure[$n=59$, $r=2/5$, $d=5$]{
\label{simfig_n59_r2b5}
\begin{tabular}{ccc}
\epsfig{file = 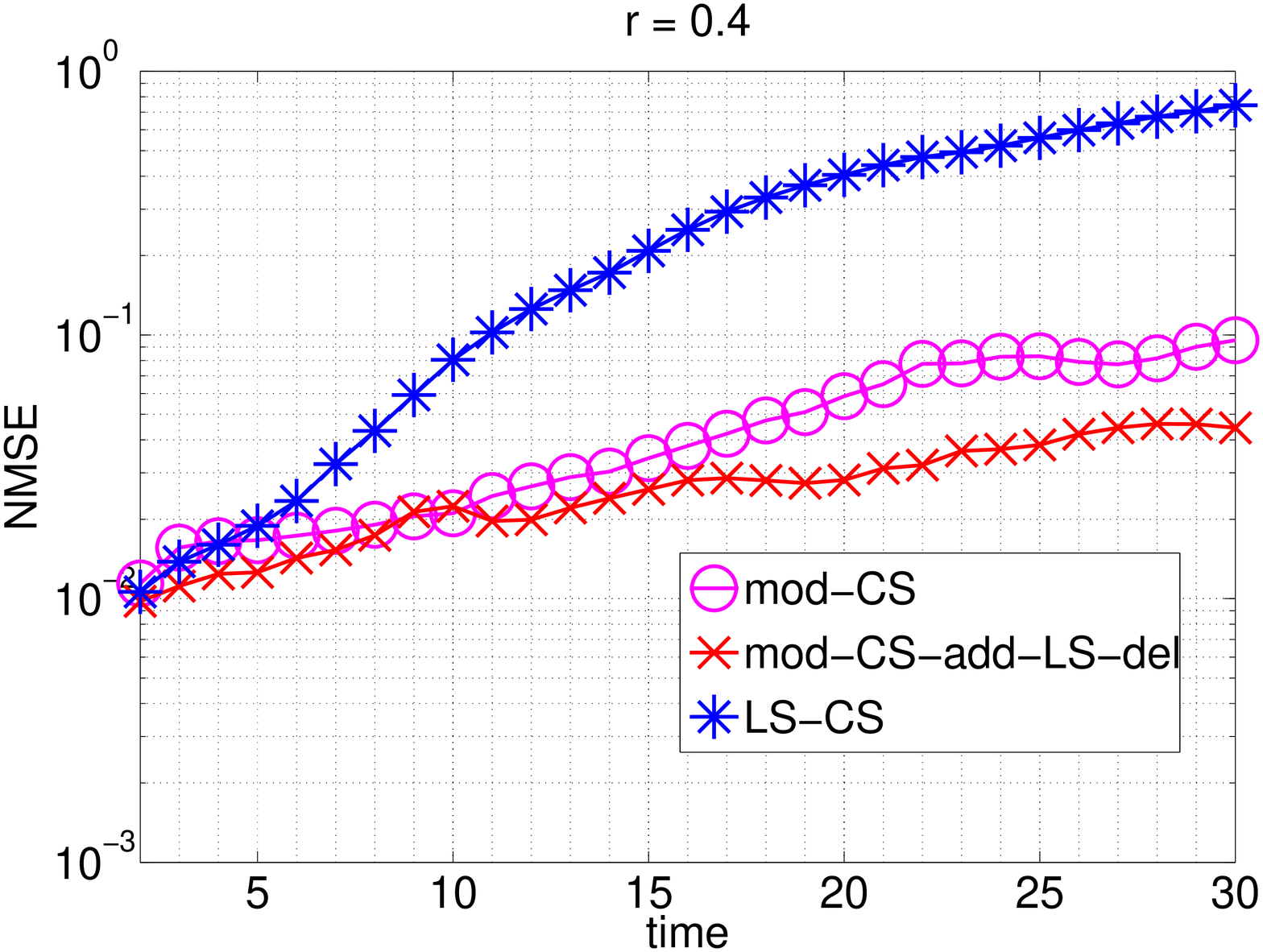, width=5.5cm} &
\epsfig{file = 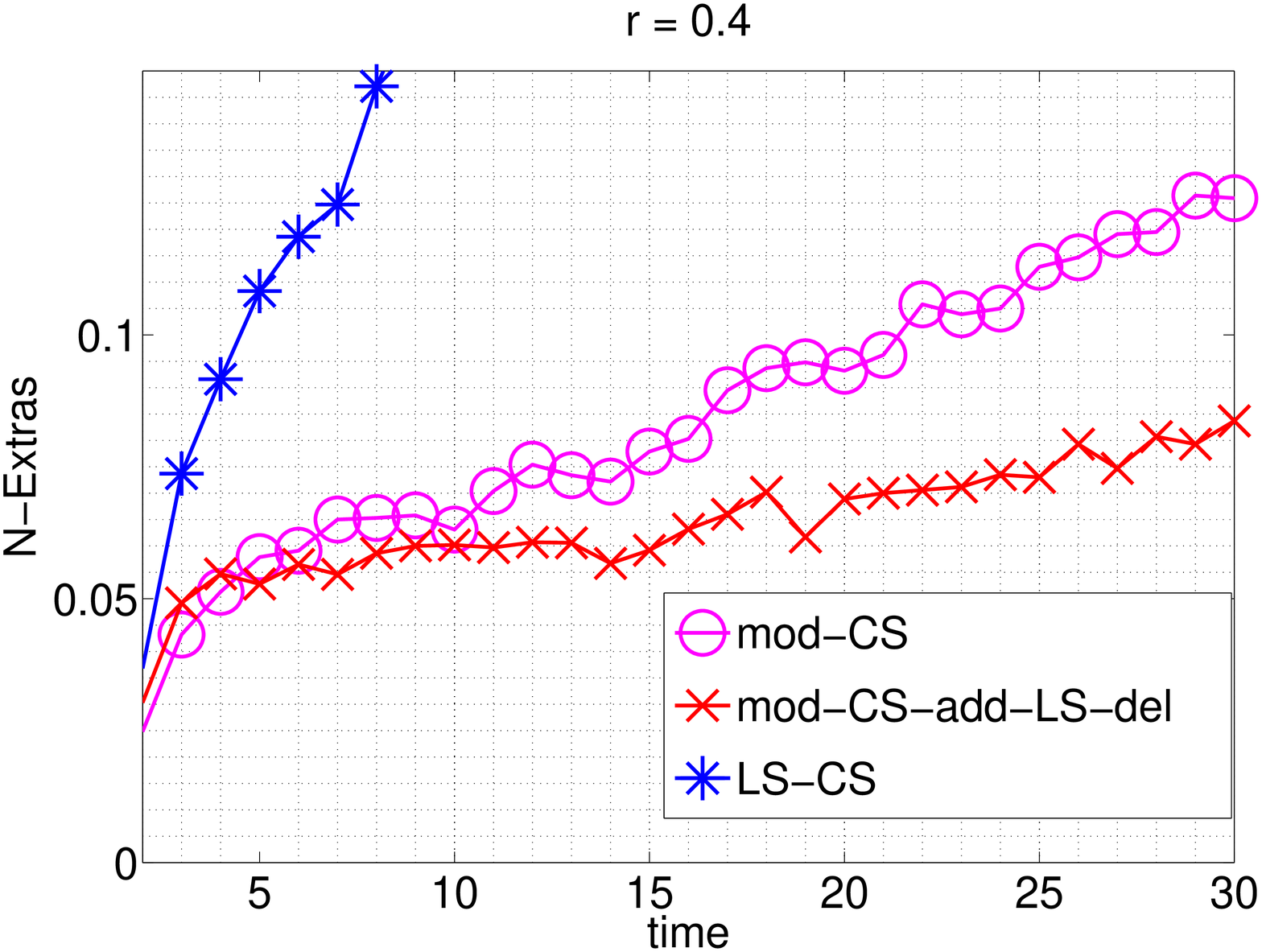, width=5.5cm} &
\epsfig{file = 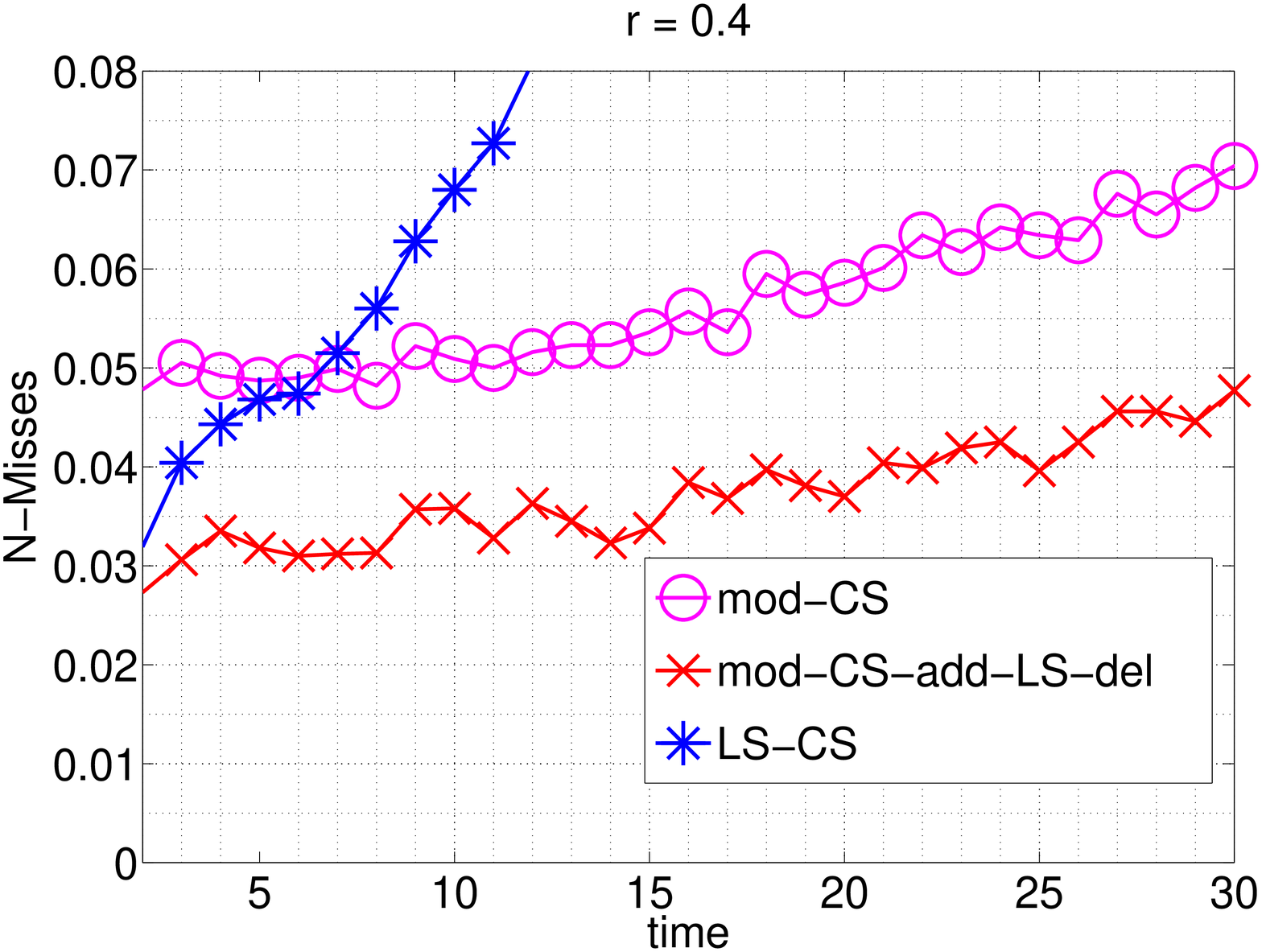, width=5.5cm}
\end{tabular}
}
}
\vspace{-0.1in}
\caption{\small{Normalized MSE (NMSE), normalized number of extras and normalized number of misses over time for modified-CS (mod-CS), modified-CS with add-LS-del (mod-CS-add-LS-del), LS-CS and simple CS. In all cases, NMSE for simple CS was more than 20\% (plotted only in (a) and (b)). We cannot use a logarithmic y-axis for plotting support errors since in some cases the errors are exactly zero.
}}
\vspace{-0.15in}
\label{fig1}
\end{figure*}



%
%
%

\bibliographystyle{IEEEbib}
\bibliography{U:/Proposals/Oct10/tipnewpfmt_kfcsfullpap}

\end{document}